\documentclass[11pt]{article}
\usepackage{amssymb}
\usepackage[mathscr]{eucal}
\usepackage{epic}
\usepackage{eepic}
\newtheorem{prop}{Proposition}
\newtheorem{remark}{Remark}
\newtheorem{defin}{Definition}

\def\EXP{\mbox{{\large\bf e}}}
\newcommand{\nc}{\newcommand}
\nc{\ds}{\displaystyle}
\nc{\qop}{\mathbf{q}}   \nc{\pop}{\mathbf{p}}   \nc{\vop}{\mathbf{v}}
\nc{\uop}{\mathbf{u}}   \nc{\wop}{\mathbf{w}}   \nc{\Weyl}{\mathfrak{W}}
\nc{\sop}{\mathbf{s}}   \nc{\xop}{\mathbf{x}}   \nc{\zop}{\mathbf{z}}
\nc{\A}{\mathcal{A}}    \nc{\R}{\mathbf{R}}     \nc{\Rer}{\,\mathbb{R}}
\nc{\Rop}{\mathcal{R}}  \nc{\Ropf}{\mathcal{R}^{(f)}}\nc{\Oc}{\mathcal{O}}
\nc{\beq}{\begin{equation}}          \nc{\eeq}{\end{equation}}
\nc{\bea}{\begin{eqnarray}}          \nc{\hx}{\hspace{3mm}}
\nc{\eea}{\end{eqnarray}}            \nc{\bdm}{\begin{displaymath}}
\nc{\ny}{\nonumber}        \nc{\om}{\omega}     \nc{\sig}{\sigma}
\nc{\hs}{\hspace{1cm}}     \nc{\hq}{\hspace{4mm}}
\nc{\ka}{\kappa}   \nc{\si}{\sigma}   \nc{\ra}{\rightarrow}  \nc{\Lam}{\Lambda}
\nc{\lk}{\left(}   \nc{\rk}{\right)}   \nc{\Rb}{\right]}  \nc{\T}{\Theta}
\nc{\lb}{\left\{}  \nc{\rb}{\right\}}  \nc{\Lb}{\left[}
\nc{\V}{\mathbf{v}\,+\,\mathbf{I}}
\nc{\vj}{\mathbf{v}}     \nc{\I}{\mathbf{I}}   \nc{\wf}{\,\mathfrak{w}}
\nc{\hal}{{\textstyle\frac{1}{2}}}       \nc{\rN}{\right)^{\!N}\hn}
\nc{\rg}{\rangle}  \nc{\Lg}{\langle}     \nc{\Rmapf}{\mathcal{R}^{(f)}}

\def\>{\rangle} \def\<{\langle}
\def\sk#1{\scriptstyle{#1}}\def\skk#1{\scriptscriptstyle{#1}}
\begin{document}
\begin{center}
{\LARGE\bf The modified tetrahedron equation and its solutions}
\\[1cm]
{\large{\bf G. von Gehlen}$^{a,}$\footnote{e-mail: gehlen@th.physik.uni-bonn.de;
   $\:^2$pakuliak@thsun1.jinr.ru;$\;\;^3$sergeev@thsun1.jinr.ru},
   $\;\;${\bf S. Pakuliak}$^{b,2},\;\;\;\;${\bf S. Sergeev}$^{b,c,3}$}\\[6mm]
$^a${\it Physikalisches Institut der Universit\"at Bonn,
Nussallee 12, D-53115 Bonn, Germany}\\[1mm]
$^b${\it Bogoliubov Laboratory of Theoretical Physics,
Joint Institute for Nuclear Research, Dubna 141980, Moscow region, Russia}
\\[1mm] $^c${\it Max-Planck-Institut f\"ur Mathematik, Vivatsgasse 7,
D-53111 Bonn, Germany}\\
\vspace*{5mm}
\phantom{Received:\hspace{1mm} \today}
\end{center}
\begin{abstract}
A large class of 3-dimensional integrable lattice spin models is constructed.
The starting point is an invertible canonical mapping operator
$\Rop_{1,2,3}$ in
the space of a triple Weyl algebra. $\Rop_{1,2,3}$ is derived postulating a
current branching principle together with a Baxter Z-invariance.
The tetrahedron equation
for $\Rop_{1,2,3}$ follows without further calculation. If the Weyl parameter
is taken to be a root of unity, $\Rop_{1,2,3}$ decomposes into a matrix
conjugation operator $\R_{1,2,3}$ and a c-number functional mapping
$\Ropf_{1,2,3}$. The operator $\R_{1,2,3}$ satisfies a modified tetrahedron
equation (MTE) in which the "rapidities" are solutions of a classical
integrable Hirota-type equations. $\R_{1,2,3}$ can be represented in terms of
the Bazhanov-Baxter Fermat curve cyclic functions, or alternatively in terms of
Gauss functions. The paper summarizes several recent publications on the
subject.
\\[3mm]
{\it AMS subject classes:}$\;\;$ 82B23,~70H06;~~~{\it keywords:}$\;$ 3D
integrable models

\end{abstract}


\section*{Introduction}
The subject of constructing and solving two-dimensional integrable lattice
models is a well-established branch of mathematical physics. It has
boosted important developments like conformal quantum field theory,
quantum groups and the quantum inverse scattering method. But the analogous
three-dimensional integrable lattice models have received little
attention: the few known cases are complicated and direct physical
applications seem to be remote. Many tools which are powerful for
dealing with two-dimensional lattice models seem to have little relevance for
the higher-dimensional cases. However, the Yang-Baxter equation can easily be
generalized to tetrahedron- and D-simplex equations \cite{bazh-stro-D,mai-nij}
which still guarantee the existence of commuting families of transfer matrices,
and concepts from the quantum inverse scattering method like L-operators can be
generalized to dimensions $D>2$ \cite{s-qem}. Since many statistical systems
require a truly three-dimensional treatment, exploring the possibilities of
constructing and solving 3D-integrable models should not be useless.

The first solution to the tetrahedron equation (TE) was found by Zamolodchikov
in 1981 \cite{Zamo}, and the partition function of this $N=2$-state model
was calculated by Baxter \cite{bax-partf} using only symmetry properties, some
factorization and commutativity. Later, Bazhanov and Baxter \cite{Bazh-Bax}
showed how to generalize the Zamolodchikov model to $N>2$, using the
interaction-round-a-cube language (IRC-ZBB-model). Then, Sergeev, Mangazeev and
Stroganov found the vertex-formulation of the ZBB-model \cite{SMS}.
These models share the feature the ZBB-Boltzmann weights are not positive
definite, but at least in the thermodynamic limit the partition function
becomes real. Baxter and Forrester \cite{bx-forr-crit} have investigated
whether the integrable ZBB-model may describe phase transitions. They modify the
ZBB-model introducing a symmetry breaking parameter (destroying the
integrability). Calculating the corresponding order parameter by variational
and numerical methods, they find good evidence that in the integrable
case the model is just at criticality (reminiscent of what happens in the
2D Potts model).

There have been several attempts to construct physically more
interesting 3D integrable models, avoiding the too restrictive Zamolodchikov TE,
using "Modified TEs" \cite{BMS,Kore-relax-TE,s-kirchhoff,s-symplectic}.
The notion of a Modified TE and the way to construct an integrable
model from its solution has been first suggested by
Mangazeev and Stroganov in \cite{manstr}.

A different starting point to find a large class of integrabel 3D-models
has been used by Sergeev \cite{s-qem,s-kiev}. First an automorphism of a
triple Weyl algebra is constructed, which satisfies a Kirchhoff-like auxiliary
linear problem and a Baxter Z-invariance principle. This invertible canonical
automorphism implies the TE, so this approach avoids the need of a tedious
proof of the TE \cite{KKS-func-te}.
If the Weyl parameter is taken to be the $N$-$th$ primitive root of unity, a
very interesting mechanism noticed before in
\cite{br-unpublished,fk-qd,br-qd,BBR-sine-gor,s-qem,pak-serg-rtoda}
 becomes operative:
The mapping operator becomes a matrix conjugation (which is the quantum
operator of an euclidian time evolution, or the Boltzmann weight) which depends
on c-number parameters, which satisfy a set of Hirota-type classical integrable
equations of motion. So we have a quantum integrable system which is
parameterized by solutions of a classical integrable system.

The different solutions of the classical integrable equations define various
3D-integrable quantum models. This framework contains the ZBB-model which is
obtained choosing the trivial classical solution. The general solution to the
Hirota-type equations for given global boundary conditions can formally be
written down in terms of high-genus theta functions, using the Fay-identity
\cite{krich,Shi,Mul,KWZ}.

What is the use of these results? It is clear that the construction of a
versatile integrable 3D-statistical model with nice physical properties
and its analytic solution is not yet round the corner, but the mathematical
structures which such a model should incorporate become more clear.
Whether this class of models is sufficiently wide to describe 3D phase
transitions, will be seen. Another application which can be made right now
is to use these 3D models in an asymmetric geometry and to reduce them to
new 2D integrable models \cite{s-rma}. Since Fermat curves are involved, these
will be useful relatives of the chiral Potts model and of
the relativistic Toda model \cite{pak-serg-rtoda}. The large variety of
parameterizations which the MTE allows (in contrast to the Yang-Baxter
equation) should help to overcome the serious parameterization problems
which have prevented progress to finding more analytic results for the chiral
Potts model.

We shall not try to touch all interesting aspects of this
program. We concentrate on explaining the construction of the basic
triple Weyl automorphism and how it gives rise to the modified tetrahedron
equation (MTE) and the classical Hirota-type equations of motion. We focus on
the local properties and do not discuss boundary conditions and large systems.
For the reader's convenience, we sketch some details of the
derivations even if these are rather straightforward. The interested reader
may consult references \cite{s-qem,pak-serg-spectr,gps} which give more
details on recent results on 3D integrable models satisfying MTEs.
\begin{figure}[ht]
\renewcommand{\dashlinestretch}{100}
\unitlength=0.07pt
\thicklines
\begin{picture}(5600,3380)(610,778)
\shade\path(2648, 2222)(2493, 2732)(3462, 1721)(2648, 2222)
\shade\path(4277, 2776)(3462, 3277)(4431, 2266)(4277, 2776)
\dashline{45}(800, 1272)(2016, 1685)(3462, 1338)
\dashline{45}(2016, 2847)(2016, 1685)(2016, 2847)
\dashline{45}(4678, 4074)(4678, 2912)
\dashline{45}(6124, 2565)(4678, 2912)(3462, 2499)
\path(2246, 924)(2246, 2086)
\path(3462, 2499)(2016, 2847)(800, 2433)(2246, 2086)(3462, 2499)
\path(3462, 2499)(3462, 1338)(2246, 924)(800, 1272)(800, 2433)
\path(3462, 3660)(4908, 3313)
\path(6124, 3726)(6124, 2565)(4908, 2151)
\path(4908, 3313)(4908, 2151)(3462, 2499)(3462, 3660)(4678, 4074)
             (6124, 3726)(4908, 3313)
\put(2548,2080){$\mathfrak{w}_1$} \put(2493,2782){$\mathfrak{w}_3$}
\put(3505,1721){$\mathfrak{w}_2$}
\put(4200,2870){$\mathfrak{w}_1'$}\put(3500,3277){$\mathfrak{w}_2'$}
\put(4471,2290){$\mathfrak{w}_3'$}
\put(3300,2585){$A$} \put(3300,3650){$C$} \put(6050,2410){$H$}
\put(2210,2140){$P$} \put(4860,2000){$B$} \put(4860,3380){$D$}
\put(4600,2730){$E$} \put(4630,4130){$F$} \put(6050,3790){$G$}
\put(1960,1520){$K$} \put(1930,2910){$R$} \put(3480,1260){$L$}
\put(2200,780){$M$} \put(630,1150){$N$} \put(630,2400){$Q$}
\blacken\path(3430,3560)(3462,3660)(3494,3560)(3430,3560)  
\blacken\path(4785,2145)(4908,2151)(4800,2210)(4785,2145)  
\blacken\path(4595,2855)(4678,2912)(4550,2905)(4595,2855)  
\end{picture}
\caption{\footnotesize
Two cubes of the basic lattice intersected by the auxiliary plane (shaded) in
two different positions: first passing through
$\wf_1,\,\wf_2,\,\wf_3$ and second
through $\wf_1',\,\wf_2',\,\wf_3'$. The second position is obtained from the
first by moving the auxiliary plane parallel through the vertex $A$. The Weyl
variables $\wf_i,\,\wf_i'$ live on the links of the basic lattice.}
\label{triangle}\end{figure}
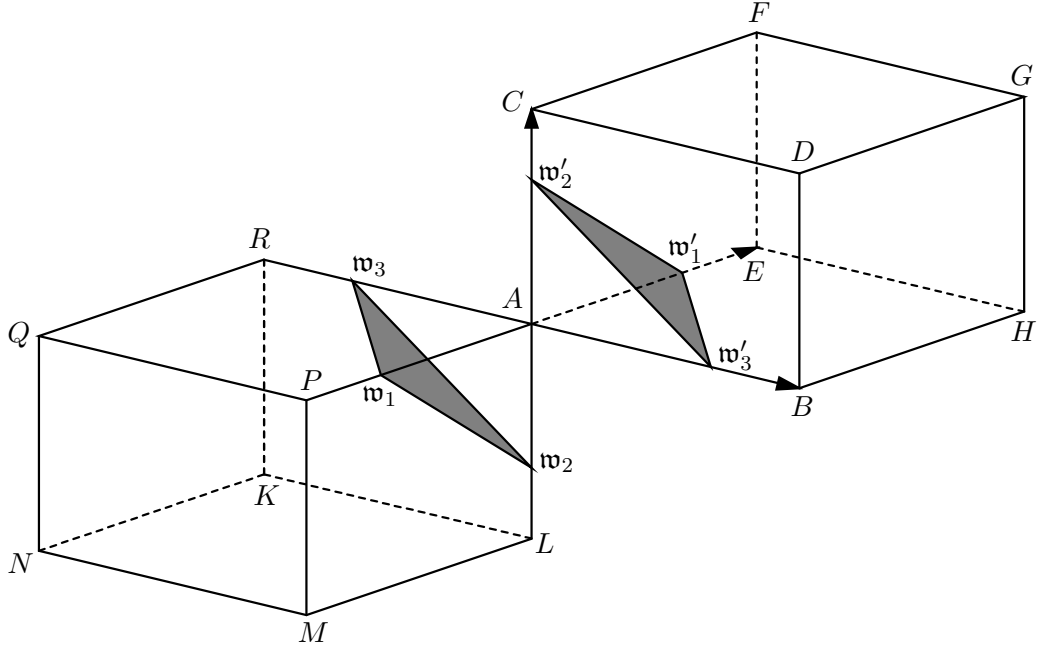
\section{The canonial invertible mapping $\Rop_{123}$}
The models to be discussed in this paper are defined on a 3-dimensional
oriented lattice, which can be imagined as a (possibly distorted) cubic
oriented lattice.
We use the vertex language: the dynamic variables $\wf_i$ live on the oriented
links $i$ of the lattice and the Boltzmann weights are associated to the
vertices.
We assume that the lattice is non-degenerate so that at all vertices exactly
three lines meet. The Boltzmann weights can be considered as operators
mapping the three dynamic variables on the incoming links to the three
variables on the outgoing links.
%
\begin{figure}[h]
\setlength{\unitlength}{0.019mm}
{\renewcommand{\dashlinestretch}{50}
\begin{picture}(7920,5000)(0,650)
%
\path(4297,5187)(7897,5187)(6322,3612)(2722,3612)(4297,5187)
\put(3850,4130){\makebox(0,0)[lb]{$\mathfrak{w}'_2$}}
\put(4360,4790){\makebox(0,0)[lb]{$\mathfrak{w}'_1$}}
\put(6330,4782){\makebox(0,0)[lb]{$\mathfrak{w}'_3$}}
\path(4297,4737)(7297,4737)                      
\blacken\path(7177,4707)(7297,4737)(7177,4767)(7177,4707)       
\path(3747,3737)(4972,4962)                      
\blacken\path(4908.360,4855.934)(4972,4962)(4865.934,4898.36)
(4908.360,4855.934)
\path(3400,3840)(7372,4962)                      
\blacken\path(7264.859,4900.188)(7372,4962)(7248.375,4957.88)
(7264.859,4900.188)
\thicklines\path(4747,4737)(4432,5500) \put(4200,5500){$E$}
\path(4022,4012)(3171,4600) \put(2970,4600){$C$}
\path(6647,4737)(7184,5500) \put(6950,5500){$B$}
\dashline{80}(5848,3612)(6647,4737)     
\blacken\path(6594.580,4613.932)(6620,4737)(6538.127,4634.255)(6594.58,4613.932)
\dashline{80}(5192,3612)(4747,4737)     
\blacken\path(4808.245,4629.533)(4747,4737)(4750.467,4613.355)(4808.245,4629.53)
\dashline{90}(4600,3612)(4022,4012)     
\blacken\path(4146.161,3961.789)(4022,4012)(4097.337,3926.915)(4146.161,3961.79)
\put(5390,2960){\makebox(0,0)[lb]{$\bullet$}}
%
\path(6097,1462)(5192,3612)         
\path(4340,1462)(5848,3612)         
\path(6760,2120)(4600,3612)         
%
\path(22,2262)(2722,2262)
\blacken\path(2482,2202)(2722,2262)(2482,2322)(2482,2202)
\path(1372,912)(1372,3612)
\blacken\path(1432,3372)(1372,3612)(1312,3372)(1432,3372)\thinlines
\path(1597,2080)(1597,1670)(2240,1670)(2240,2080)(1597,2080)(1410,2226)
\blacken\path(1530,2177)(1410,2226)(1480,2140)(1530,2177)
%
\dottedline{80}(3950,1617)(2727,1937)
\blacken\path(2850.691,1937.711)(2727,1937)(2836.474,1879.42)
(2850.691,1937.711)
%
\path(4297,2587)(7897,2587)(6322,1012)(2722,1012)(4297,2587)
\put(4310,1260){$\mathfrak{w}_3$}
\put(6220,1260){$\mathfrak{w}_1$}
\put(6700,1880){$\mathfrak{w}_2$}
\path(3600,1462)(6547,1462)                    
\blacken\path(6427,1432)(6547,1462)(6427,1492)(6427,1432)
\path(5872,1237)(7097,2462)                    
\blacken\path(7033.360,2355.934)(7097,2462)(6990.934,2398.36)
(7033.360,2355.934)
\path(3622,1260)(7297,2267)                    
\blacken\path(7189.667,2205.52)(7297,2267)(7173.363,2263.264)
(7189.667,2205.52)
\thicklines\dashline{80}(6282.5,1012)(6080,1492) \path(6411,700)(6282.5,1012)
\put(6520,700){\makebox(0,0)[lb]{$P$}} \put(3940,700){\makebox(0,0)[lb]{$R$}}
\put(7540,1650){\makebox(0,0)[lb]{$L$}}\put(5590,3050){\makebox(0,0)[lb]{$A$}}
\dashline{80}(7151,1850)(6760,2120) \path(7512,1600)(7151,1850)
\dashline{90}(4023,1012)(4359,1492) \path(3803,700)(4023,1012)\thinlines
%
\path(4297,1732)(4237.119,1714.392)  (4200.020,1704.083)  (4160.950,1692.921)
    (4122.012,1681.019)    (4085.306,1668.490)    (4027.000,1642.000)
\path(4027,1642) (3992.474,1616.589)    (3953.222,1581.242)
    (3917.110,1542.525)    (3892.000,1507.000)
\path(3892,1507) (3869.200,1460.073)    (3847.866,1401.258)
    (3837.349,1339.063)    (3847.000,1282.000)
\path(3847,1282) (3887.714,1229.806)    (3948.190,1190.635)
    (4014.321,1163.397)    (4072.000,1147.000)
\path(4072,1147) (4128.683,1138.773)  (4195.659,1136.761) (4231.881,1137.733)
    (4269.326,1139.836)    (4307.544,1142.928)    (4346.084,1146.869)
    (4384.496,1151.517)    (4422.331,1156.730)    (4459.138,1162.369)
    (4494.467,1168.292)    (4558.890,1180.425)    (4612.000,1192.000)
\path(4612,1192) (4663.518,1207.673)
    (4726.761,1231.529)    (4788.874,1258.120)    (4837,1282)
\path(4837,1282) (4879.455,1308.020)
    (4930.997,1343.406)    (4980.542,1381.840)    (5017,1417)
\path(5017,1417) (5045.042,1452.197)
    (5075.826,1497.756)    (5099.697,1547.937)    (5107,1597)
\path(5107,1597) (5091.355,1650.383)
    (5058.726,1702.259)    (5016.484,1746.505)    (4972,1777)
\path(4972,1777) (4909.336,1791.563)  (4872.449,1792.170)  (4834.176,1789.945)
    (4796.257,1786.102)    (4760.428,1781.854)    (4702,1777)
\path(4702,1777) (4642.667,1778.429)  (4606.048,1779.679)  (4567.412,1780.810)
    (4528.725,1781.465)    (4491.951,1781.286)    (4432,1777)
\path(4432,1777) (4363.656,1756.337)    (4297,1732)
%
\path(1372,912) (1442.930,919.968)
    (1508.875,927.767)    (1570.080,935.448)    (1626.794,943.060)
    (1679.264,950.652)    (1727.736,958.274)    (1772.459,965.974)
    (1813.679,973.804)    (1851.643,981.811)    (1886.599,990.046)
    (1948.474,1007.396)  (2001.282,1026.249)    (2047,1047)
\path(2047,1047)    (2081.547,1065.839) (2118.788,1088.434) (2158.205,1114.405)
    (2199.276,1143.375)    (2241.482,1174.963)    (2284.304,1208.792)
    (2327.221,1244.482)    (2369.714,1281.656)    (2411.262,1319.935)
    (2451.347,1358.939)    (2489.448,1398.290)    (2525.045,1437.610)
    (2557.618,1476.519)    (2586.649,1514.640)    (2611.616,1551.593)(2632,1587)
\path(2632,1587)    (2659.338,1654.703) (2670.663,1694.173) (2680.551,1736.602)
    (2689.099,1781.402)    (2696.402,1827.982)    (2702.556,1875.753)
    (2707.656,1924.125)    (2711.799,1972.509)    (2715.079,2020.315)
    (2717.592,2066.953)    (2719.434,2111.835)    (2720.700,2154.370)
    (2721.486,2193.969)    (2722.000,2262.000)
\path(2722,2262)    (2721.311,2330.056) (2720.335,2369.682) (2718.835,2412.251)
    (2716.734,2457.169)    (2713.955,2503.846)    (2710.423,2551.688)
    (2706.059,2600.104)    (2700.787,2648.500)    (2694.530,2696.286)
    (2687.211,2742.868)    (2678.754,2787.655)    (2669.081,2830.054)
    (2658.115,2869.472)    (2632.000,2937.000)
\path(2632,2937)    (2612.009,2973.931) (2587.649,3012.973)(2559.381,3053.663)
    (2527.669,3095.539)    (2492.976,3138.138)    (2455.764,3180.996)
    (2416.498,3223.650)    (2375.639,3265.639)    (2333.650,3306.498)
    (2290.996,3345.764)    (2248.138,3382.976)    (2205.539,3417.669)
    (2163.663,3449.381)    (2122.973,3477.649)    (2083.931,3502.009)(2047,3522)
\path(2047,3522)    (1979.472,3548.115) (1940.054,3559.081) (1897.655,3568.754)
    (1852.868,3577.211)    (1806.286,3584.530)    (1758.500,3590.787)
    (1710.104,3596.059)    (1661.688,3600.423)    (1613.846,3603.955)
    (1567.169,3606.734)    (1522.251,3608.835)    (1479.682,3610.335)
    (1440.056,3611.311)    (1372.000,3612.000)
\path(1372,3612)    (1303.944,3611.311) (1264.318,3610.335) (1221.749,3608.835)
    (1176.831,3606.734)  (1130.154,3603.955)   (1082.312,3600.423)
    (1033.896,3596.059)   (985.500,3590.787)    (937.714,3584.530)
    (891.132,3577.211)    (846.345,3568.754)    (803.946,3559.081)
    (764.528,3548.115)    (697.000,3522.000)
\path(697,3522) (660.071,3502.008)    (621.030,3477.646)    (580.340,3449.377)
    (538.465,3417.665)    (495.867,3382.971)    (453.009,3345.760)
    (410.354,3306.494)    (368.365,3265.635)    (327.506,3223.647)
    (288.238,3180.993)    (251.026,3138.136)    (216.333,3095.538)
    (184.620,3053.662)    (156.352,3012.972)    (131.991,2973.931) (112,2937)
\path(112,2937) (85.885,2869.472)    (74.919,2830.054)    (65.246,2787.655)
    (56.789,2742.868)    (49.470,2696.286)    (43.213,2648.500)
    (37.941,2600.104)    (33.577,2551.688)    (30.045,2503.846)
    (27.266,2457.169)    (25.165,2412.251)    (23.665,2369.682)
    (22.689,2330.056)  (22,2262)
\path(22,2262)  (22.689,2193.944)
    (23.665,2154.318)    (25.165,2111.749)    (27.266,2066.831)
    (30.045,2020.154)    (33.577,1972.312)    (37.941,1923.896)
    (43.213,1875.500)    (49.470,1827.714)    (56.789,1781.132)
        (65.246,1736.345)    (74.919,1693.946)    (85.885,1654.528)   (112,1587)
\path(112,1587) (131.992,1550.071) (156.354,1511.030) (184.623,1470.340)
    (216.335,1428.465)    (251.029,1385.867)    (288.240,1343.009)
    (327.506,1300.354)    (368.365,1258.365)    (410.353,1217.506)
    (453.007,1178.238)    (495.864,1141.026)    (538.462,1106.333)
    (580.338,1074.620)    (621.028,1046.352)    (660.069,1021.991) (697,1002)
\path(697,1002) (743.831,981.831)  (797.433,964.737) (859.785,950.456)
    (894.860,944.287)    (932.864,938.723)    (974.044,933.729)
   (1018.647,929.274)   (1066.921,925.324)   (1119.113,921.847)
   (1175.469,918.808)   (1236.238,916.176)   (1301.666,913.918)   (1372,912)
\put(1325,2210){\makebox(0,0)[lb]{$\bullet$}}
\put(5590,3050){\makebox(0,0)[lb]{$A$}}
\put(1460,1660){\makebox(0,0)[lb]
  {$\begin{array}{c}\;\;\mathbf{u}_3\;\mathbf{w}_3\\[-1.5mm]\ka_3\end{array}$}}
\put( 610,2700){\makebox(0,0)[lb]{$|\phi_3\rg$}}
\put(1530,2700){\makebox(0,0)[lb]{$q^{1/2}\uop_3|\phi_3\rg$}}
\put( 500,1680){\makebox(0,0)[lb]{$\wop_3|\phi_3\rg$}}
\put(1430,1330){\makebox(0,0)[lb]{$\ka_3\uop_3\wop_3|\phi_3\rg$}}
\end{picture}}
\caption{\footnotesize Right hand side: magnified view of the vicinity of point
$A$ in Fig.\ref{triangle}.
Left: magnified view of the auxiliary plane in the vicinity of $\wf_3$ showing
the four sectors formed by the lines $\:\overline{\wf_3\,\wf_2}\:$ and
$\:\overline{\wf_3\,\wf_1}\,.:$ We imagine a current $|\phi_3\rg$ flowing out
of the intersection point into the four 2-dimensional sectors with a
distribution steered by the Weyl variable $\wf_3$ and the coupling $\ka_3$.}
\label{Sergm}
\end{figure}
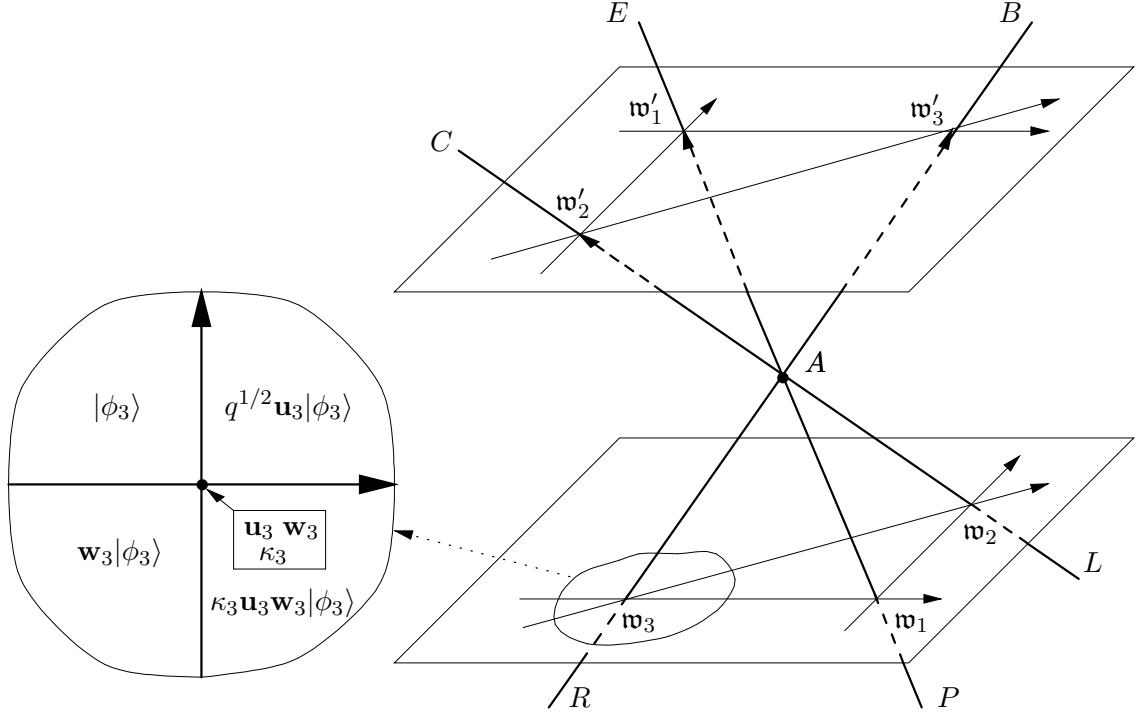
Fig.\ref{triangle} shows a perspective view of two adjacent cubes of the
lattice (to keep the picture simple, we do not show the other vertices and
lines). Only the dynamic variables on the links surrounding the vertex $A$ are
indicated. The links (or edges) of the lattice arise from the intersection
of two planes each, the vertices from the intersection of three planes.
We introduce auxiliary planes (shaded in Fig.\ref{triangle})
and locate the dynamic variables at the vertices of the lines
where the lattice planes intersect the auxiliary planes. The mapping of the
variables $\wf_i$ onto the $\wf_i'$ can be regarded as moving the auxiliary
plane through the vertex $A$. This way we may interpret the 3-dimensional
statistical model as the euclidian time evolution of the 2-dimensional lattice
which arises when we move the auxiliary plane.

The next step is to specify the mapping, or, equivalently, the Boltzmann
weights. We assume that the dynamic variables $\wf_i$ are elements of the
ultralocal Weyl algebra $\wf_i\:\in\:\Weyl^{\otimes\Delta}$,
defined as the tensor product of $\Delta$ copies of Weyl pairs ($\Delta$ is
the number of vertices in the auxiliary plane):
\beq  \Weyl^{\otimes\Delta}\;=\;\{
 \uop_i\:=\: 1 \otimes 1 \otimes ... \,\otimes
\underbrace{\uop}_{{i-th\atop place}}\otimes\, ...\;\:, \;\;\;
\wop_i\:=\: 1 \otimes 1 \otimes ...\, \otimes
\underbrace{\wop}_{{i-th\atop place}}\otimes\, ...\;|\;i=1,\ldots,\Delta\:\}
\label{weyl}  \end{equation}   \vspace*{-5mm}  with
\beq\uop_i\,\wop_j\:=\:q^{\delta_{i,j}}\wop_j\,\uop_i,\hs\;q\:\in\:{\mathbb{C}}.
\label{Weyl}  \end{equation}
In order to determine the mapping uniquely, we postulate a linear current
branching principle (a kind of Kirchhoff law), together with a Baxter
Z-invariance. Consider Fig.\ref{Sergm}. On the right hand side we show again
the vicinity of the vertex $A$ of Fig.\ref{triangle}: on top and on the bottom
there are the two auxiliary planes. The three links of the basic lattice which
connect the $\wf_i$ and the $\wf_i'$, intersect in $A$.
We now postulate currents $\;|\phi_i\rg\;$ flowing out of the vertices into the
surrounding four sectors of the auxiliary plane. The distribution of the current
$\;|\phi_i\rg\;$ flowing out of $\:\wf_i\:$ into the sectors is steered by
the value of the variable $\:\wf_i\:$ and one coupling constant $\:\ka_i\:$,
as shown in the left hand picture of Fig.\ref{Sergm}:
The vertex $i$ sends the current $\:|\phi_i\rg\:$ into the sector left to the
arrows, into the sector between the two outgoing arrows it sends
$\;q^{1/2}\uop_i|\phi_i\rg\:$, to the right of the arrows
$\;\wop_i|\phi_i\rg\:$, and between the incoming arrows
$\;\ka_i\uop_i\wop_i|\phi_i\rg\:$.
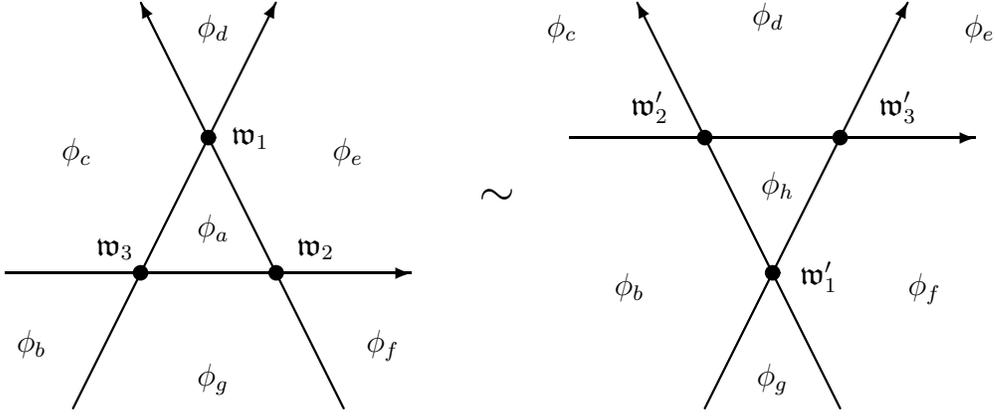
\begin{figure}[h]
\setlength{\unitlength}{0.3mm}
\thicklines
\begin{picture}(450,200)
\put(00,00){\begin{picture}(200,180)
{\large \put( 10,70){\vector(1,0){180}} \put(40,10){\vector(1,2){90}}
        \put(160,10){\vector(-1,2){90}} \put(70,70){\circle*{6}}
                \put( 48, 78){$\wf_3$}   \put(130,70){\circle*{6}}
                \put(137, 78){$\wf_2$}   \put(100,130){\circle*{6}}
                \put(108,127){$\wf_1$}   \put(95,86){$\phi_a$}
                \put( 95,175){$\phi_d$}     \put(95,20){$\phi_g$}
                \put( 35,120){$\phi_c$}     \put(155,120){$\phi_e$}
                \put( 15, 35){$\phi_b$}     \put(170, 35){$\phi_f$}}
\end{picture}}
\put(250,00){\begin{picture}(200,200)
{\large \put( 10,130){\vector(1,0){180}} \put(70,10){\vector(1,2){90}}
        \put(130, 10){\vector(-1,2){90}} \put(70,130 ){\circle*{6}}
                \put( 35,140){$\wf'_2$}          \put(130,130){\circle*{6}}
                \put(145,140){$\wf'_3$}          \put(100, 70){\circle*{6}}
                \put(110, 65){$\wf'_1$}          \put( 95,105){$\phi_h$}
                \put( 91,180){$\phi_d$}          \put( 93, 20){$\phi_g$}
                \put(  0,175){$\phi_c$}          \put(185,175){$\phi_e$}
                \put( 30, 60){$\phi_b$}          \put(160, 60){$\phi_f$}}
\end{picture}}
\put(210,90){\begin{picture}(30,50)
\put(10,10){{\LARGE $\sim $}}\end{picture}}\end{picture}
\caption{\footnotesize The triangles on the lower and upper auxiliary plane.}
\label{fig-YBE}   \end{figure}
The total current in an internal sector is required to be zero and the total
current reaching an external sector should be independent of the internal
structure. In Fig.\ref{fig-YBE} we show the lower and upper auxiliary planes of
the previous two Figures. According to the rules just stated, the total
currents
flowing into the internal sectors $\;|\phi_a\rg\:$ and $\;|\phi_h\rg\;$ must
both vanish. $\,|\phi_a\rg\;$ receives
contributions from $\:|\phi_1\rg,\;|\phi_2\rg\:$ and $\;|\phi_3\rg$.
Since both structures differ just by a shift of one line through one vertex,
there is a unique correspondence between the respective sectors.
So these rules give:\bea
|\phi_a\rg &=&\wop_1|\phi_1\rg\:+\:|\phi_2\rg\:+\:q^{1/2}\uop_3|\phi_3\rg\;
                      =\;0 \ny\\
|\phi_b\rg &=&|\phi_1'\rg\:+\:\wop_2'|\phi_2'\rg\;=\;\wop_3|\phi_3\rg \ny\\
|\phi_c\rg &=&|\phi_2'\rg\:\;=\;|\phi_1\rg\:+\:|\phi_3\rg    \ny\\
|\phi_d\rg &=&|\phi_3'\rg\:+\:q^{1/2}\uop_2'|\phi_2'\rg\;
                                            =\;q^{1/2}\uop_1|\phi_1\rg \ny\\
|\phi_e\rg &=&q^{1/2}\uop_3'|\phi_3'\rg\;=\;\ka_1 \uop_1 \wop_1|\phi_1\rg\:
                                          +\:q^{1/2}\uop_2|\phi_2 \rg \ny \\
|\phi_f\rg &=&\ka_1\uop_1'\wop_1'|\phi_1'\rg\:+\:\ka_3\uop_3'\wop_3'|\phi_3'\rg
                             \;=\;\ka_2\uop_2\wop_2|\phi_2\rg \ny\\
|\phi_g\rg &=&\wop_1'|\phi_1'\rg\;=\;\wop_2|\phi_2\rg\:
                      +\:\ka_3\uop_3\wop_3|\phi_3\rg \ny\\
|\phi_h\rg &=&q^{1/2}\uop_1'|\phi_1'\rg\:+\:\ka_2\uop_2'\wop_2'|\phi_2'\rg\:
                 +\:\wop_3'|\phi_3'\rg\;=\;0    \label{kirch}   \eea
These eight equations determine the $\uop_i'$ and $\wop_i'$ uniquely in terms
of the $\uop_i$ and $\wop_i$ and vice versa (we give details of the calculation
in the Appendix). The resulting rational transformation can be written as:
\beq  \ds\left.\begin{array}{lll}
\ds\wop_1'\;=\;\ds\wop_2\;\Lam_3,&\ds\wop_2'\;=\;\ds\Lam_3^{-1}\wop_1\,,\;\;&
\ds\wop_3'\;=\;\ds\Lam_{2}^{-1}\uop_1^{-1}\;, \\[3mm]
\ds\uop_1'\;\:=\;\,\ds\Lam_{2}^{-1}\wop_3^{-1}\,,\;\;\:&
\ds\uop_2'\;=\;\ds\Lam_{1}^{-1}\uop_3\,,\;\;&
\ds\uop_3'\;=\;\ds\uop_2^{}\:\Lam_1\,, \end{array}\right.\label{Rma}\eeq
where
\beq \left.\begin{array}{ccl}\ds\Lam_1& = & \ds\uop_1^{-1}\uop_3\,-\,
q^{1/2}\,\uop_1^{-1}\wop_1\,+\,\ka_1\;\wop_1\uop_2^{-1}\,,\\[3mm]
 \ds\Lam_2& = & \ds{\ka_1\over\ka_2}\,\uop_2^{-1}\wop_3^{-1}\,+\,
{\ka_3\over\ka_2}\,\uop_1^{-1}\wop_2^{-1}\,-\,{\ka_1\,\ka_3\over q^{1/2}\:\ka_2}
\;\uop_2^{-1}\wop_2^{-1}\;,\\[4mm]
\ds\Lam_{3}^{} & = & \ds \wop_1\wop_3^{-1}\,-\,q^{1/2}\;\uop_3\wop_3^{-1}
\,+\,\ka_3\;\wop_2^{-1}\uop_3\;.\\ \end{array}\right.  \label{Lam}\eeq
In (\ref{Rma}) the order of the factors on the right hand side is
irrelevant since e.g. $\Lambda_1$ contains only operators which commute with
both $\uop_2$ and $\uop_3$, in particular, it contains no $\wop_2,\:\wop_3$.
\\[1mm]
The six eqs.(\ref{Rma}) are written with only three operators $\Lambda_i$
because the mapping conserves three centers:
\beq   \wop_1'\wop_2'=\wop_2\wop_1;\hs \uop_3'\uop_2'=\uop_2\uop_3;
    \hs {\wop_3'}^{-1}\uop_1'=\uop_1\wop_3^{-1}. \label{cuwi}   \eeq
This rational mapping conserves the Weyl structure, i.e. it is canonical:
from (\ref{Weyl}) it follows that
\beq \uop_i'\,\wop_j'\:=\:q^{\delta_{i,j}}\wop_j'\,\uop_i',\end{equation}
as may be verified by explicit use of (\ref{Rma}) and (\ref{Lam}).
\begin{defin}
For any rational function $\;\Phi\;$ of the $\;\uop_1,\ldots,\wop_3,$
the relations (\ref{Rma}), (\ref{Lam}) define the invertible and
canonical mapping $\;\Rop_{1,2,3}\;$
\beq \lk\Rop_{1,2,3}\circ\Phi\rk
(\uop_1,\wop_1,\uop_2,\ldots,\wop_3)\stackrel{def}{=}
\Phi(\uop_1',\wop_1',\uop_2',\ldots,\wop_3').  \label{Rop} \end{equation}
$\:\Rop_{1,2,3}\:$ is an
automorphism of $\:\Weyl^{\otimes\Delta}$ and conserves the three centers
(\ref{cuwi}).
\end{defin}
The derivation of (\ref{Rma}), (\ref{Lam}) from (\ref{kirch}) given in the
Appendix does not use that
$\uop_i,\:\wop_j$ are Weyl operators. However, the canonical property
of $\;\Rop_{1,2,3}\;$ is obtained for ultralocal Weyl dynamic variables only.
The inverse transformation has a similar form and we summarize it giving
$\Lambda_i^{-1}$ in terms of the primed variables:
\beq \left.\begin{array}{ccl}
 \ds\Lam_1^{-1}& = & \ds{\ka_1\over\ka_2}\,\uop_1'{\uop_3'}^{-1}\,-\,q^{1/2}
{\ka_3\over\ka_2}\,\uop_1'{\wop_1'}^{-1}\,+\,\ka_3\;\uop_2'{\wop_1'}^{-1}\;,
\\[3mm]
\ds\Lam_2^{-1} & = & \ds \uop_2'\wop_3'\,-\,q^{-1/2}\:\ka_2\:\uop_2'\wop_2'
\,+\,\uop_1'\wop_2'\;,\\[3mm]
\ds\Lam_3^{-1}& = & \ds{\ka_3\over\ka_2}\,{\wop_1'}^{-1}\wop_3'\,-\,q^{1/2}
{\ka_1\over\ka_2}\,{\uop_3'}^{-1}\wop_3'\,+\,\ka_1\;\wop_2'{\uop_3'}^{-1}\;.
 \end{array}\right.   \label{Lami} \eeq

There is a way dual to (\ref{kirch}) to formulate the current
branching principle as "Linear problem". This is more convenient
if one is dealing with a large lattice with definite boundary
conditions. The resulting mapping is the same
(\ref{Rma}),~(\ref{Lam}), we give this alternative derivation in
the Appendix. We refer to \cite{pak-serg-spectr,svan} for a
detailed exposition of the linear problem equations for large
cubic lattices.

\section{Weyl parameter $q$ a root of unity: decomposition of $\Rop$ into a
matrix conjugation and a functional mapping.}
Up to now, $q$ was in general position. In all the following we fix $q$
to be a primitive root of unity:
\beq q\;=\:\omega\:\stackrel{def}{=}\;e^{2\pi i/N},\hs
          N\in \mathbb{Z},\hs N\ge 2.\eeq
Then $\;\uop^N\:$ and $\:\wop^N\:$ are centers of the Weyl algebra and we can
represent the canonical pair $(\uop_i,\,\wop_i)$ by its action on a cyclic
basis as unitary $N\times N$ matrices multiplied by complex parameters
$\,u_i,\:w_i.$ The algebra $\Weyl^{\otimes\Delta}$ will then
be represented in a $N^\Delta$-dimensional product space, see (\ref{weyl}).
Because of the ultra-locality, the Weyl operators $(\uop_i,\:\wop_i)\in \wf_i$
will be represented trivially in all spaces $j\neq i$, acting non-trivially
only in the space with label $i$, where (now omitting the index $i$):
\bea \uop\;&=&u\,\xop\;;\hs\hs\wop\;=\;w\,\zop\;;\ny \\[2mm]
 |\,\sigma\rg&\equiv&|\,\sigma\;\;mod\;N\>\;;\hq
\langle\,\sigma\,|\,\sigma' \rg\;=\;\delta_{\sigma,\sigma'};\hq
\xop\,|\,\sigma\rg\;=\;|\,\sigma\rg\,\om^\sigma\;;\hq
\zop\,|\,\sigma\rg\;=\;|\,\sigma+1\rg\:.\hs \label{XZdef}  \eea
In this representation the centers are represented by numbers:
\beq   \uop_i^N\;=\;u_i^N\;,\hs\hs  \wop_i^N\;=\;w_i^N\;.\label{rootu}
\end{equation}
We shall see that for $q$ a root of unity, the rather complicated looking
transformation (\ref{Rma}),(\ref{Lam}) can be represented in a rather
simple way.
\subsection{The functional mapping $\;\Rop_{1,2,3}^{(f)}$}\label{funm}
For $q$ a root of unity, not only the $N$-th powers of the Weyl variables
are numbers, but also the $N$-$th$ powers of the operators $\:\Lambda_i$ will
be numbers. This is because in calculating the $N$-th powers the cross terms
drop out, e.g. $\:(a\,\uop\,+\,b\,\wop)^N\:=\:(a\,u)^N\,+\,(b\,w)^N$ due to
$\sum_{j=0}^{N-1}\omega^j=0$, etc. We get
\beq\label{lambda-N}
\ds\left.\begin{array}{ccl} \ds\Lam_1^N & = & \ds
u_1^{-N}\,u_3^{N}\;+\;u_1^{-N}\,w_1^{N}\;+\;\ka_1^N\;w_1^{N}\,u_2^{-N}\;,\\[2mm]
\ds\Lam_2^N & = & \ds
{\ka_1^N\over\ka_2^N}\;u_2^{-N}\, w_3^{-N}\;+\;{\ka_3^N\over\ka_2^N}\;
u_1^{-N}\, w_2^{-N}\;+\;{\ka_1^N\;\ka_3^N\over\ka_2^N}\;u_2^{-N}\,w_2^{-N}\;,
\\[4mm]\ds\Lam_3^N & = & \ds w_1^{N}\, w_3^{-N}\;+\;u_3^{N}\, w_3^{-N}\;+\;
\ka_3^N\;w_2^{-N}\,u_3^{N}\,.\end{array}\right.
\end{equation}
So (\ref{Rma}) implies an analogous purely functional mapping of the Weyl
centers:
\beq   \ds\left.\begin{array}{lll}
\ds{w_1'}^N\;=\;\Lam_3^N\:w_2^N,&\ds{w_2'}^N\;=\;\Lam_3^{-N}\:w_1^N, &
\ds{w_3'}^N\;=\;\Lam_2^{-N}\: u_1^{-N}, \\[3mm]
\ds{u_1'}^N\;=\;\Lam_2^{-N}\:w_3^{-N},&\ds{u_2'}^N\;=\;\Lam_1^{-N}\:u_3^N,&
\ds{u_3'}^N\;=\;\Lam_1^{N}\:u_2^N.
\end{array}\right.\label{funcN}   \end{equation}
Later we shall need not only the mapping of the centers (\ref{funcN}) but
also the mapping of the $u_i,\;w_i$ itself onto the $u_i',\;w_i'$. For this we
have to take $N$-$th$ roots and get a freedom to choose phases. If we
demand the number products $w_1w_2,\;u_2u_3,\;u_1w_3^{-1}$ to be centers of
this mapping, then there are three free phases.
 \begin{defin}
The functional counterpart of the mapping $\Rop_{1,2,3}$
is the mapping $\:\Rop_{1,2,3}^{(f)}$, acting on the space
of functions of the parameters $u_j,\:w_j\;\:(j=1,2,3)$
\beq \lk\Rop_{1,2,3}^{(f)}\circ\phi\rk
(u_1^{},w_1^{},u_2^{},w_2^{},u_3^{},w_3^{})\;
\stackrel{def}{=}\;\phi(u_1',w_1',u_2',w_2',u_3',w_3')\;,
   \label{fctm} \end{equation}
where the primed variables are functions of the unprimed ones,
defined via
\beq  {u_1'}^N\,=\,{\uop_1'}^N\:,\hq
{w_1'}^N\,=\,{\wop_1'}^N\:,\hq \mbox{etc.,}\end{equation}
such that $\;\; w_1'w_2'\,=\,w_1w_2\;,\;\;\;u_2'u_3'\,=\,u_2u_3\;,\;\;\;
u_1'{w_3'}^{-1}\,=\,u_1w_3^{-1}\,.$
The three free phases of the $N$-$th$ roots are extra discrete parameters of
$\Rop_{1,2,3}^{(f)}$.
\end{defin}
In obvious generalization:
\begin{remark}
To any rational automorphism of the ultra-local Weyl algebra
$\Weyl^{\otimes\Delta}$ (\ref{weyl}) there is a rational mapping in the space
of the $N$-th powers of the parameters of the representation: $u_j^N,\;w_j^N$.
\end{remark}
\subsection{Matrix structure of the mapping $\Rop_{1,2,3}$}\label{rco}
We now look at the matrix form of $\Rop_{1,2,3}$ in the representation
(\ref{XZdef}). We separate the numerical factors from the constant
diagonal resp. cyclic raising matrices $\xop_i,\;\zop_i$.
Understanding that the matrices $\xop_i, \;\wop_i$ act trivially in the
spaces with $j\neq i$, from (\ref{Rma}) we immediately get the following
complicated looking $N^3\times N^3$-matrix equations:
\bea  \lk\xop_1'\rk^{-1}&=&\frac{\ka_1u'_1}{\ka_2u_2}\xop_2^{-1}
  +\frac{\ka_3u'_1w_3}{\ka_2u_1w_2}\xop_1^{-1}\zop_2^{-1}\zop_3
  -\om^{1/2}\frac{\ka_1\ka_3u'_1w_3}
  {\ka_2u_2w_2}\xop_2^{-1}\zop_2^{-1}\zop_3, \ny\\[2mm]
\zop_1'&=&\frac{w_2w_1}{w'_1w_3} \zop_1 \zop_2 \zop_3^{-1}
  -\om^{1/2}\frac{w_2u_3}{w'_1w_3}\zop_2 \xop_3 \zop_3^{-1}
  +\frac{\ka_3}{w'_1}\xop_3, \ny \\[2mm]
\lk\xop_2'\rk^{-1}&=&\frac{u'_2}{u_1}\xop_1^{-1}-\om^{1/2}
\frac{w_1u'_2}{u_1u_3}\xop_1^{-1}\zop_1 \xop_3^{-1}
  +\frac{\ka_2w_1u'_2}{u_2u_3}\zop_1\xop_2^{-1}\xop_3^{-1},\ny \\[2mm]
\lk\zop_2'\rk^{-1}&=&\frac{w'_2}{w_3}\zop_3^{-1}
-\om^{1/2}\frac{w'_2u_3}{w_1w_3}\zop_1^{-1}
   \xop_3 \zop_3^{-1}+\frac{\ka_3w'_2u_3}{w_1w_2}\zop_1^{-1}
\zop_2^{-1}\xop_3,
    \ny\\[2mm]
\xop_3'&=&\frac{u_2u_3}{u_1u'_3} \xop_1^{-1}\xop_2\xop_3
-\om^{1/2} \frac{w_1u_2}{u_1u'_3}\xop_1^{-1}\zop_1 \xop_2
  +\frac{\ka_2w_1}{u'_3}\zop_1,\ny\\[2mm]
\lk\zop_3'\rk^{-1}&=& \frac{\ka_1u_1w'_3}{\ka_2u_2w_3}\xop_1
\xop_2^{-1}\zop_3^{-1}
  +\frac{\ka_3w'_3}{\ka_2w_2}\zop_2^{-1}
  -\om^{1/2}\frac{\ka_1\ka_3u_1w'_3}{\ka_2u_2w_2}\xop_1
\xop_2^{-1}\zop_2^{-1}.\label{matx}   \eea
However, eqs.(\ref{matx}) can be rewritten in a compact way because there
exists a (up to a scalar factor) unique $N^3\times N^3$ conjugation
matrix $\;\R_{1,2,3}\,$, such that for $i=1,\,2,\,3$
\beq  \xop_i'\;=\;\R_{1,2,3}\:\xop_i\:\R_{1,2,3}^{-1}\:,\hs
\zop_i'\;=\;\R_{1,2,3}\:\zop_i\:\R_{1,2,3}^{-1}\:,\label{Rdef} \end{equation}
and there is a closed formula for the matrix elements of $\:\R_{1,2,3}\:$ which
later will be given explicitly. Let us write the matrix elements of
$\:\R_{1,2,3}$ in our representation (\ref{XZdef}) as
\beq \langle i_1,i_2,i_3|\R_{1,2,3}|j_1,j_2,j_3\,\rangle\;
\stackrel{{\rm def}}{=}\; R_{i_1,i_2,i_3}^{j_1,j_2,j_3}.\label{rel}
\end{equation}
Then e.g. the from the equation for $\xop_3'$ in (\ref{matx})
we see that we have to find $\:R_{i_1,i_2,i_3}^{j_1,j_2,j_3}\:$ such that
it satisfies the recursion
\beq \lk \om^{j_3}-\frac{u_2'}{u_1}\om^{-i_1+i_2+i_3}\rk
R_{i_1,i_2,i_3}^{j_1,j_2,j_3}+\om^{-1/2+i_2-i_1}\frac{w_1u_2}{u_1u_3'}
   R_{i_1,i_2,i_3}^{j_1,j_2-1,j_3}
   -\frac{\ka_2w_1}{u_3'}\,R_{i_1,i_2,i_3}^{j_1,j_2,j_3-1}\,=\,0\,,
   \label{examx} \end{equation}
together with five other similar relations from (\ref{matx}).
The indices are cyclic mod $N$.
\\[2mm]
Now, going back to (\ref{Rop}), we remark that in our representation also any
rational function $\Phi$ of the $\,\uop_1,\ldots,\wop_3\,$ is represented by a
$N^3\times N^3$ matrix with entries which are rational functions of
$\,u_1,\ldots,w_3\,$. The mapping $\;\Rop_{1,2,3}\;$ transforms the matrix
entries by $\;\Rop_{1,2,3}^{(f)}$ and conjugates the matrix by $\:\R_{1,2,3}\:$.
So we get the
\begin{remark} $\;\;$
At $\,q\,$ a root of unity, the mapping $\Rop_{1,2,3}$ can be represented
as the superposi\-tion of the pure functional mapping $\:\Rop^{(f)}_{1,2,3}\:$
and the finite dimensional similarity transformation $\:\R_{1,2,3}\:$:
\beq  \Rop_{1,2,3}\:\circ\:\Phi\;=\;
\R_{1,2,3}^{}\;\lk\Rop^{(f)}_{1,2,3}\circ\Phi\rk\;\R_{1,2,3}^{-1}\:.
\label{frfr}\end{equation}
\end{remark}
This remarkable property arising in models where the quantum variables
are elements of a Weyl algebra at root of unity, has been pointed out in
\cite{br-unpublished} and in \cite{Fa-Vo}, \cite{fk-qd}.
\section{Tetrahedron equation and modified tetrahedron
equation}\label{tetrah}
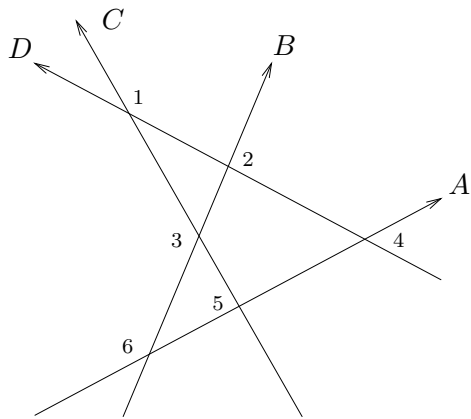
\begin{figure}[ht]
\setlength{\unitlength}{0.016mm}
\begin{center}
{\renewcommand{\dashlinestretch}{30}
\begin{picture}(3670,3450)(0,660)
\path(959,675)(2191,3613)
\path(2172,3491)(2191,3613)(2116,3513)
\path(3600,1813)(225,3613)
\path(345,3583)(225,3613)(316.8,3530)
\path(2445,675)(570,3965)
\path(654.502,3874.670)(570.000,3965.000)(602.052,3845.532)
\path(225,688)(3600,2488)
\path(3508.235,2405.059)(3600.000,2488.000)(3480.000,2458.000)
\put(3660,2520){\makebox(0,0)[lb]{$A$}}
\put(2200,3680){\makebox(0,0)[lb]{$B$}}
\put( 780,3900){\makebox(0,0)[lb]{$C$}}
\put(   0,3658){\makebox(0,0)[lb]{$D$}}
\put(1035,3270){\makebox(0,0)[lb]{$\sk{1}$}}
\put(1950,2758){\makebox(0,0)[lb]{$\sk{2}$}}
\put(1355,2083){\makebox(0,0)[lb]{$\sk{3}$}}
\put(3200,2083){\makebox(0,0)[lb]{$\sk{4}$}}
\put(1700,1563){\makebox(0,0)[lb]{$\sk{5}$}}
\put( 950,1200){\makebox(0,0)[lb]{$\sk{6}$}}
\end{picture}}
\caption{\footnotesize Quadrangle in the auxiliary plane formed by the directed
intersection lines of 4 oriented lattice planes.}\label{quadro}
\end{center}
\end{figure}

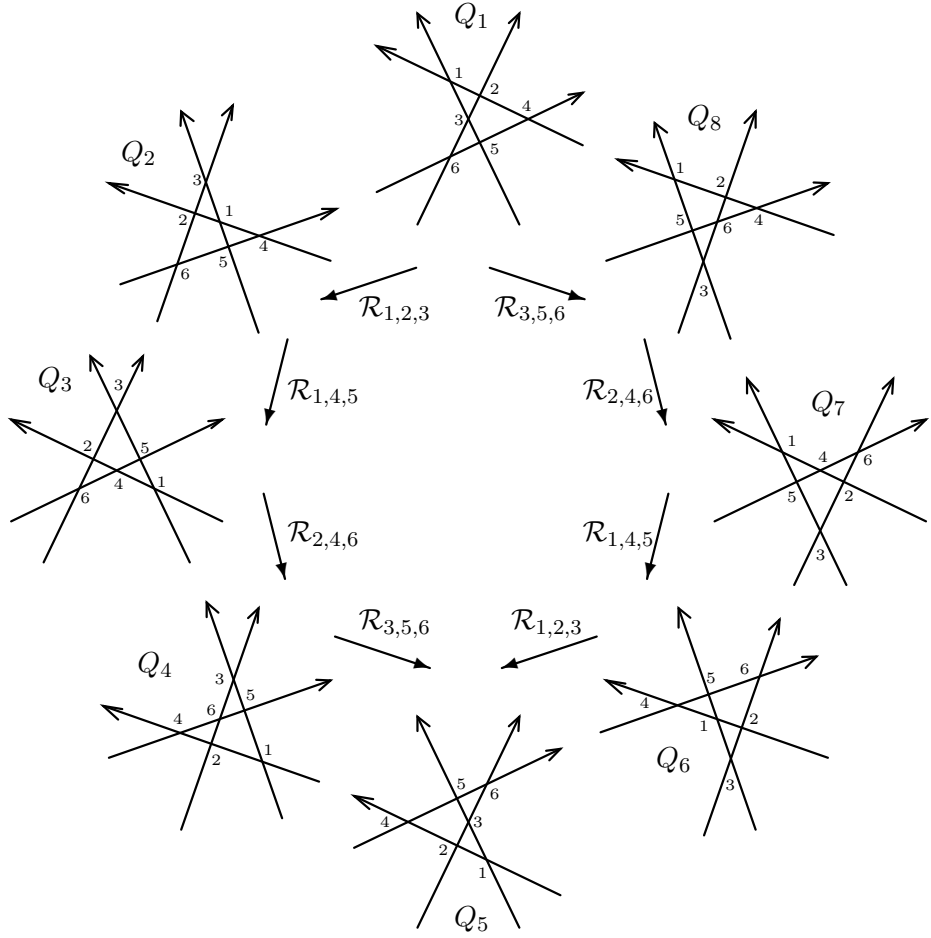
\begin{figure}[ht]
\begin{center}
\unitlength=0.09pt
\thicklines
\begin{picture}(3800,3870)(1520,570)
\put(4900,2750){$Q_7$}
\path(5048, 2018)(4631, 2884)\path(4631,2823)(4631, 2884)(4680,2833)
\path(4829, 2018)(5246, 2884)\path(5202,2843)(5246, 2884)(5242,2822)
\path(4495, 2285)(5383, 2713)\path(5323,2705)(5383, 2713)(5336,2666)
\path(4495, 2713)(5383, 2285)\path(4558,2660)(4495, 2713)(4575,2700)
\put(4800,2600){$\skk1$} \put(4930,2535){$\skk4$} \put(5120,2520){$\skk6$}
\put(4800,2365){$\skk5$} \put(5040,2370){$\skk2$} \put(4920,2120){$\skk3$}
\put(4250,1250){$Q_6$}
\path(4768, 1871)(4450,  964)\path(4724,1830)(4768, 1871)(4764,1810)
\path(4923, 1716)(4016, 1398)\path(4863,1714)(4923, 1716)(4877,1668)
\path(4971, 1291)(4040, 1616)\path(4103,1560)(4040, 1616)(4120,1602)
\path(4668,  988)(4343, 1919)\path(4343,1838)(4343, 1919)(4393,1880)
\put(4600,1640){$\skk6$} \put(4460,1605){$\skk5$} \put(4180,1500){$\skk4$}
\put(4430,1400){$\skk1$} \put(4540,1155){$\skk3$} \put(4640,1430){$\skk2$}
\put(3400,580){$Q_5$}
\path(2980,  911)(3846, 1328) \path(3786,1320)(3846, 1328)(3800,1280)
\path(3674, 1464)(3246,  576) \path(3630,1423)(3674, 1464)(3670,1403)
\path(2980, 1130)(3846,  713) \path(3043,1075)(2980, 1130)(3060,1116)
\path(3246, 1464)(3674,  576) \path(3246,1403)(3246, 1464)(3296,1423)
\put(3500,785){$\skk1$} \put(3335,880){$\skk2$} \put(3090,1000){$\skk4$}
\put(3480,1000){$\skk3$} \put(3550,1115){$\skk6$} \put(3410,1160){$\skk5$}
\put(2070,1650){$Q_4$}
\path(2832, 1191)(1925, 1509) \path(1988, 1452)(1925, 1509)(2005,1496)
\path(2677, 1036)(2360, 1943) \path(2360, 1882)(2360, 1943)(2410,1901)
\path(2252,  988)(2578, 1919) \path(2534, 1878)(2578, 1919)(2574,1858)
\path(1950, 1291)(2880, 1616) \path(2820, 1618)(2880, 1616)(2834,1574)
\put(2395,1600){$\skk3$} \put(2380,1260){$\skk2$} \put(2600,1300){$\skk1$}
\put(2220,1430){$\skk4$} \put(2360,1470){$\skk6$} \put(2525,1530){$\skk5$}
\put(1650,2850){$Q_3$}
\path(2426, 2285)(1538, 2713)\path(1600, 2657)(1538, 2713)(1618,2700)
\path(2289, 2113)(1872, 2979)\path(1872, 2918)(1872, 2979)(1922,2938)
\path(1675, 2113)(2092, 2979)\path(2048, 2938)(2092, 2979)(2088,2918)
\path(1538, 2285)(2426, 2713)\path(2366, 2706)(2426, 2713)(2380,2677)
\put(1970,2835){$\skk3$} \put(2080,2570){$\skk5$} \put(2150,2430){$\skk1$}
\put(1840,2580){$\skk2$} \put(1830,2360){$\skk6$} \put(1970,2420){$\skk4$}
\put(2000,3800){$Q_2$}
\path(1998, 3282)(2905, 3599) \path(2845, 3599)(2905, 3599)(2840,3555)
\path(2153, 3127)(2470, 4034) \path(2424, 3993)(2470, 4034)(2464,3973)
\path(2578, 3079)(2252, 4009) \path(2250, 3948)(2252, 4009)(2305,3968)
\path(2880, 3381)(1950, 3707) \path(2013, 3665)(1950, 3707)(2030,3700)
\put(2300,3700){$\skk3$} \put(2435,3570){$\skk1$} \put(2580,3420){$\skk4$}
\put(2240,3530){$\skk2$} \put(2250,3305){$\skk6$} \put(2410,3355){$\skk5$}
\put(3400,4380){$Q_1$}
\path(3075, 3670)(3941, 4087) \path(3880, 4079)(3941, 4087)(3895,4040)
\path(3246, 3533)(3674, 4421) \path(3630, 4380)(3674, 4421)(3670,4360)
\path(3674, 3533)(3246, 4421) \path(3248, 4360)(3246, 4421)(3296,4380)
\path(3941, 3867)(3075, 4284) \path(3132, 4227)(3075, 4284)(3155,4270)
\put(3400,4150){$\skk1$} \put(3550,4080){$\skk2$} \put(3685,4010){$\skk4$}
\put(3400,3950){$\skk3$} \put(3385,3750){$\skk6$} \put(3550,3830){$\skk5$}
\put(4380,3950){$Q_8$}
\path(4088, 3806)(4995, 3489) \path(4150, 3760)(4088, 3806)(4169,3800)
\path(4561, 3054)(4243, 3961) \path(4240, 3900)(4243, 3961)(4295,3920)
\path(4343, 3079)(4668, 4009) \path(4620, 3960)(4668, 4009)(4668,3940)
\path(4040, 3381)(4971, 3707) \path(4910, 3660)(4971, 3707)(4910,3710)
\put(4330,3750){$\skk1$} \put(4500,3690){$\skk2$} \put(4660,3530){$\skk4$}
\put(4330,3530){$\skk5$} \put(4430,3230){$\skk3$} \put(4530,3495){$\skk6$}
\put(3240,3350){\vector(-3,-1){400}}\put(3550,3350){\vector(3,-1){400}}
\put(2700,3050){\vector(-1,-4){90}} \put(4200,3050){\vector(1,-4){90}}
\put(2600,2400){\vector(1,-4){90}}  \put(4300,2400){\vector(-1,-4){90}}
\put(2900,1800){\vector(3,-1){400}} \put(4000,1800){\vector(-3,-1){400}}
\put(3000,3150){${\mathcal R}_{1,2,3}$}\put(3570,3150){${\mathcal R}_{3,5,6}$}
\put(2700,2800){${\mathcal R}_{1,4,5}$}\put(3940,2800){${\mathcal R}_{2,4,6}$}
\put(2700,2200){${\mathcal R}_{2,4,6}$}\put(3940,2200){${\mathcal R}_{1,4,5}$}
\put(3000,1830){${\mathcal R}_{3,5,6}$}\put(3640,1830){${\mathcal R}_{1,2,3}$}
\end{picture}
\end{center}
\caption{\footnotesize{Graphical image of the two equivalent ways of
transforming the four-line-graph ("quadrilateral") $Q_1$ into graph
$Q_5$, which leads to tetrahedron equation. Observe that each graph
contains only two triangles which can be transformed by a mapping
$\Rop$: In graph $Q_1$ either the line 124 can be moved
downward through the point 3 (leading to graph $Q_2$), or the line 456
can be moved upward through point 3 (leading to $Q_8$). Both the left
hand and right hand sequences of four transformations lead to the same
graph $Q_5.$}} \label{tet-graph}
\end{figure}

We postpone for a moment the discussion of the explicit form of
$\R_{1,2,3}$ and of the possible form of $\Rop_{1,2,3}^{(f)}$.
We shall first see that, once we have constructed the invertible canonical
mapping of the triple Weyl algebra $\Rop_{1,2,3}$, no calculation is needed
to conclude that $\Rop_{i,j,k}$ satisfies the Tetrahedron equation.
We want to keep the point of view taken in Fig.\ref{triangle} that the mapping
$\Rop_{1,2,3}$ comes about by moving the auxiliary plane through a vertex
of the basic lattice. For considering the single mapping $\:\Rop_{1,2,3}\:$
we selected {\it three} planes of the basic lattice and intersected these by the
auxiliary plane. Now we look at {\it four} (with no pair being parallel) lattice
planes intersecting the auxiliary plane. In the auxiliary plane this gives rise to
a figure called quadrangle shown in Fig.\ref{quadro}. A quadrangle consists of
four lines which form two triangles and one four-sided figure. There are
six crossing points.
%
%
%
Shifting the auxiliary plane within the basic lattice, we can reverse
(from $\Delta$ to $\nabla$) either of the two triangles: Shifting such that the
plane producing the line $B$ gets to the other side of point 1 (this happens
moving the auxiliary plane through the vertex formed by planes $B,C,D$) we
perform the mapping $\Rop_{1,2,3}$. Alternatively, we may shift the plane $C$
through the point 6 (moving the auxiliary plane through the vertex of $A,B,C$),
so performing the mapping $\Rop_{3,5,6}$. \\[2mm]
As shown in Fig.\ref{tet-graph}, there are two different sequences of mappings
which, starting from $Q_1$, lead to the same quadrangle $Q_5$: either
$Q_1\ra Q_2\ra Q_3\ra Q_4\ra Q_5$ or
$Q_1\ra Q_8\ra Q_7\ra Q_6\ra Q_5$.
Since the mapping $\Rop_{i,j,k}$ is invertible and canonical, we
conclude that the two corresponding products of the $\Rop$ must be the
same mapping:
\beq\label{op-te}
\ds \Rop_{123}\cdot\Rop_{145}\cdot\Rop_{246}\cdot\Rop_{356}\;\sim\;
\Rop_{356}\cdot\Rop_{246}\cdot\Rop_{145}\cdot\Rop_{123}\,,\label{tetra}
\end{equation}
i.e. that $\Rop_{i,j,k}$ satisfies the Tetrahedron equation\footnote{We use
the following notation for the superposition of two mappings
$\mathcal{A}$ and $\mathcal{B}$:
\beq \ds\lk\lk\mathcal{A}\cdot\mathcal{B}\rk\circ\Phi\rk\;
\stackrel{def}{=}\;
\lk\mathcal{A}\circ\lk\mathcal{B}\circ\Phi\rk\rk\;.
\end{equation}}.
%
%
%
(\ref{tetra}) is an operator equation in the space of the six Weyl variables.
An equivalent interpretation is that the six dynamic variables live on the six
edges of a tetrahedron, and the right hand side of (\ref{tetra}) is obtained
from the left hand side by moving one corner of the tetrahedron through the
plane defined by the other three corners.
Inserting the decomposition (\ref{frfr}) of $\Rop_{i,j,k}$ into a matrix- and
a functional part into (\ref{tetra}), we get
\bea \lefteqn{
\R_{123}\Lb\Ropf_{123}\lb\R_{145}\lk \Ropf_{145}\Lb\R_{246}\lb\Ropf_{246}
    \lk\R_{356}\lk\Ropf_{356}\circ\Phi\rk\R_{356}^{-1}\rk\rb\R_{246}^{-1}\Rb\rk
        \R_{145}^{-1}\rb\Rb\R_{123}^{-1}}\ny\\ &=&\:
\R_{356}\Lb\Ropf_{356}\lb\R_{246}\lk \Ropf_{246}\Lb\R_{145}\lb\Ropf_{145}
    \lk\R_{123}\lk\Ropf_{123}\circ\Phi\rk\R_{123}^{-1}\rk\rb\R_{145}^{-1}\Rb\rk
        \R_{246}^{-1}\rb\Rb\R_{356}^{-1}.\ny\\&&        \label{tea}
                \eea
In (\ref{tea}) we have to apply the functional transformation repeatedly to
matrices $\R_{i,j,k}$ and to $\Ropf_{i,j,k}$ and $\Phi$. Let us write
shorthand
\bea \R^{(1)}&=&\R_{123};\hs \R^{(2)}\;=\;\Ropf_{1,2,3}\circ\R_{145};\hs
\R^{(3)}\;=\;\Ropf_{1,2,3}\Ropf_{1,4,5}\circ\R_{246};\ny\\ 
\R^{(4)}&=&\Ropf_{1,2,3}\Ropf_{1,4,5}\Ropf_{2,4,6}\circ\R_{356};\hs\hs\hq\;
\R^{(5)}\;=\;\Ropf_{3,5,6}\Ropf_{2,4,6}\Ropf_{1,4,5}\circ\R_{123};\ny\\  
\R^{(6)}&=&\Ropf_{3,5,6}\Ropf_{2,4,6}\circ\R_{145};\hs
\R^{(7)}\;=\;\Ropf_{3,5,6}\circ\R_{246};\hs \R^{(8)}\;=\;\R_{356}.
   \label{teab}\eea
Then (\ref{tea}) can be rewritten as
\bea \lefteqn{
\lk\R^{(1)}\,\R^{(2)}\,\R^{(3)}\,\R^{(4)}\rk \;
\lb \Rop^{(f)}_{123}\cdot\Rop^{(f)}_{145}
\cdot\Rop^{(f)}_{246}\cdot\Rop^{(f)}_{356}\,\Phi\,\rb\;
\lk \R^{(1)}\R^{(2)}\R^{(3)}\R^{(4)}\rk^{-1}}\ny\\ \!\! &=&
\!\!\lk\R^{(8)}\,\R^{(7)}\,\R^{(6)}\,\R^{(5)}\rk \;
\lb \Rop^{(f)}_{356}\cdot\Rop^{(f)}_{246}
\cdot\Rop^{(f)}_{145}\cdot\Rop^{(f)}_{123}\,\Phi\,\rb\;
\lk \R^{(8)}\R^{(7)}\R^{(6)}\R^{(5)}\rk^{-1}.\hx \label{rff}
\eea
Since, as has been discussed in Sec. \ref{funm}, any rational automorphism
of the ultra-local Weyl algebra implies a rational mapping in the space
of the $N$-$th$ powers of the parameters of the representation,
its is a direct consequence of (\ref{op-te}) that the $\Rop^{(f)}_{i,j,k}$ of
(\ref{fctm})
solve the tetrahedron equation with the variables
$\:u_j^N,\;w_j^N,\;j=1,\ldots,6$: \\
\beq\label{fc-te} \lk \Rop^{(f)}_{123}\,\cdot\,\Rop^{(f)}_{145}\,\cdot\,
\Rop^{(f)}_{246}\,\cdot\,\Rop^{(f)}_{356}\;-\;\Rop^{(f)}_{356}\,\cdot\,
\Rop^{(f)}_{246}\,\cdot\,\Rop^{(f)}_{145}\,\cdot\,\Rop^{(f)}_{123}\rk\:
\phi(u_1^N,\ldots,w_6^N)\;=\;0.
\end{equation}
Using Maple, by a ten-line-half-minute-program one can easily check the
validity of this functional equation straight from the definitions
(\ref{funcN}), (\ref{lambda-N}).

Now the $\,\R^{j}\,$ depend not on the $u_i^N,\;w_i^N\:$ but on the
$\,u_i,\;w_i$, see (\ref{matx}). So we have to chose the phases, which are at
our disposal in defining $\,\Ropf_{i,j,k}\,$ such that it solves the
{\it functional} tetrahedron equation for rational functions of the variables
$u_j$ and $w_j$ directly. However, not all phases of the $u_j,\;w_j$ can be
chosen independently because there are various centers which should be
conserved. We shall not enter into the discussion of these phase choices here,
in \cite{gps} it is shown that 16 phases can be chosen arbitrarily.
Once a consistent choice of phases has been made, we obtain (\ref{fc-te}) in the
variables $u_j,\;w_j$ and can cancel the central functional terms in
(\ref{rff}). So we arrive at the
\begin{prop}
The {\it finite dimensional} matrices $\R$ satisfy the Modified Tetrahedron
Equation (MTE):
\beq\label{mte-general}
\ds\begin{array}{l}\ds\R_{1,2,3}
\cdot\lk\Ropf_{1,2,3}\circ\R_{1,4,5}\rk
\!\cdot\!\lk\Ropf_{1,2,3}\Ropf_{1,4,5}\circ\R_{2,4,6}\rk
\!\cdot\!\lk\Ropf_{1,2,3}\Ropf_{1,4,5}\Ropf_{2,4,6}\circ\R_{3,5,6}\rk
\\[4mm]
\sim\; \R_{3,5,6}\cdot\lk\Ropf_{3,5,6}\circ\R_{2,4,6}\rk
\!\cdot\!\lk\Ropf_{3,5,6}\Ropf_{2,4,6}\circ\R_{1,4,5}\rk
\!\cdot\!\lk\Ropf_{3,5,6}\Ropf_{2,4,6}\Ropf_{1,4,5}\circ\R_{1,2,3}\rk\,.
\end{array}\end{equation}
\end{prop}
The left and right hand sides of (\ref{mte-general})
may differ by a scalar factor, which arises
when we pass from the equivalence of the mappings to the equality
of the matrices.
So, writing all matrix indices and using again the abbreviations (\ref{teab}),
the MTE reads
\beq\label{mte-elements}\ds\begin{array}{l}\ds \sum_{j_1...j_6}\;
\lk R^{(1)}\rk_{i_1,i_2,i_3}^{j_1,j_2,j_3}\,
\lk R^{(2)}\rk_{j_1,i_4,i_5}^{k_1,j_4,j_5}\,
\lk R^{(3)}\rk_{j_2,j_4,i_6}^{k_2,k_4,j_6}\,
\lk R^{(4)}\rk_{j_3,j_5,j_6}^{k_3,k_5,k_6}\\[4mm]
\ds\hs\hq\;=\;\rho\,\sum_{j_1...j_6}\;
\lk R^{(8)}\rk_{i_3,i_5,i_6}^{j_3,j_5,j_6}\,
\lk R^{(7)}\rk_{i_2,i_4,j_6}^{j_2,j_4,k_6}\,
\lk R^{(6)}\rk_{i_1,j_4,j_5}^{j_1,k_4,k_5}\,
\lk R^{(5)}\rk_{j_1,j_2,j_3}^{k_1,k_2,k_3}\,.
\end{array}\end{equation}
Here $\rho$ is the scalar density factor. Taking determinants in
(\ref{mte-elements}) we can express the $N^3$-$th$ power of the scalar factor
as
\beq  \rho^{N^3}\;=\;
\frac{\ds \det\R^{(1)}\ \det\R^{(2)}\ \det\R^{(3)}\ \det\R^{(4)}}
{\ds \det \R^{(8)}\ \det \R^{(7)}\ \det \R^{(6)}\ \det \R^{(5)}},
\label{rrho}\end{equation}
This can be obtained from the determinant of one single matrix
$\;\R_{1,2,3}\;$ just by substituting the respective arguments.
The components of the eight $\R^{(j)}$ may be considered as Boltzmann weights
of the models. However, in general these matrix elements will not be positive.
We call eqs. (\ref{mte-general}) or (\ref{mte-elements}) {\it modified}
tetrahedron equations because in (\ref{mte-general}) the $\R^{(j)}$ depend on
several "rapidity" variables $u_1,\ldots,w_6$ which are not the same on the
left- and right hand sides of the equation, but rather are related by various
functional mappings. We shall later see that there is much freedom in choosing
$\Ropf_{1,2,3}$, accordingly leading to different models.

\section{Details on the matrix- and the functional transformations}
\subsection{Matrix part of $\;\Rop_{1,2,3}\;$ at root of unity
in terms of Fermat curve\\ cyclic functions $W_p(n)$}

In Sec.\ref{rco} we introduced the conjugation operator $\R_{1,2,3}$
and its representation $R_{i_1,i_2,i_3}^{j_1,j_2,j_3}$, which allowed to
write eqs.(\ref{matx}) in the form (\ref{Rdef}). We just have seen that the
matrix elements of the $\R_{1,2,3}$ play the role of Boltzmann weights and
solve the MTE.

In (\ref{examx}) we had remarked that $R_{i_1,i_2,i_3}^{j_1,j_2,j_3}$ has to
satisfy six recursion relations with respect to its modulo $N$ defined indices,
but have not yet seen its explicit form.\\[2mm]
We now show that the $R_{i_1,i_2,i_3}^{j_1,j_2,j_3}$ can be constructed
from Fermat curve cyclic functions $W_p(n)$ which were introduced by
Bazhanov and Baxter \cite{Bazh-Bax} for building up the Boltzmann weights of
their model. The form (\ref{RWdef}) which we shall use has first been given
in \cite{SMS} several years ago.\\[1mm]
Introduce a two component vector $p=(x,y)$ which is restricted to the Fermat
curve $x^N+y^N=1$. Then define the function $W_p(n)$ by
\beq  W_p(0)\;=\;1,\hs W_p(n)\;=\;\prod_{\nu=1}^n\;\frac{y}{1\,-\,\om^\nu\,x}
  \hq \mbox{for}\;\;n>0.               \label{W-def}     \end{equation}
$W_p(n)\:$ satisfies the obvious recursion relation
\beq  W_p(n-1)\:=\:(1\,-\,\om^n\,x)\,y^{-1}W_p(n). \label{wrec}\end{equation}
and is cyclic in $\,n\,$: $\;W_p(n+N)\;=\;W_p(n),\:$ because of
$\;\prod_{\nu=0}^{N-1}\:(1-\om^\nu\:x)\;=y^N\;.$
Observe that we need the Fermat curve restriction in order to get the cyclic
property.
In order to satisfy the {\it six} recursion relation required by (\ref{matx}),
$R_{i_1,i_2,i_3}^{j_1,j_2,j_3}$ must be taken to be a kind of cross ratio of
four functions $W_p$, depending on four Fermat points $p_1,\ldots,p_4$:
\begin{prop}
In the basis (\ref{XZdef}) the mapping relations (\ref{matx}), (\ref{Rdef}) are
solved by the matrix
\beq   R_{i_1,i_2,i_3}^{j_1,j_2,j_3}\;=\;
\delta_{i_2+i_3,j_2+j_3}\;\om^{(j_1-i_1)\,j_3}\;
{W_{p_1}(i_2-i_1)\,W_{p_2}(j_2-j_1)\over W_{p_3}(j_2-i_1)\,W_{p_4}(i_2-j_1)}\,
  \label{RWdef}   \end{equation}
where the $x$-coordinates of the four Fermat curve points are connected by
\beq  x_1\;x_2\;=\;\om\;\;x_3\;x_4\,,\label{RM}\end{equation}
and the Fermat points are given in terms of the variables
$u_j,\; w_j,\;\ka_j,\;\;j=1,2,3\;$ by
\bea \label{xy-uw}
 x_1&=&{\om^{-1/2}\over\ka_1}\,{u_2\over u_1}\;,\hq
 x_2\;=\;\om^{-1/2}\ka_2\,{u_2'\over u_1'}\;\hq
 x_3\;=\;\om^{-1}{u_2'\over u_1}\;,\hq
 x_4\;=\;\om^{-1}{\ka_2\over\ka_1}\,{u_2\over u_1'}\;,\ny\\
{y_3\over y_1}&=&\ka_1\,{w_1\over u_3'}\;,\hq
{y_4\over y_1}\;=\;\om^{-1/2}\ka_3\,{w_3\over w_2}\;,\hq
{y_3\over y_2}\;=\;{w_2'\over w_3}\;,\hq
{y_4\over y_2}\;=\;\om^{-1/2}{\ka_3\over\ka_1}\,{u_3'\over w_1'}\;,\eea
The $\;u_j',\;w_j'$ and $\;u_j,\;w_j$ are related by the functional
transformation (\ref{fctm}):
\beq u_j'=\Rop^{(f)}_{1,2,3}\circ u_j,\hs w_j'=\Rop^{(f)}_{1,2,3}\circ w_j.
          \label{us-f}\end{equation}
\end{prop}
In order to prove this proposition, we have to check the six recursion
relations implied by $(\ref{matx})$ and have to compare the Fermat point
coefficients arising from (\ref{wrec}) with the Weyl coefficients in
(\ref{matx}). We give one example:\\
$\:R_{i_1,i_2,i_3}^{j_1,j_2,j_3}\:$ satisfies
$$ R^{j_1,j_2,j_3}_{i_1,i_2+1,i_3-1}=
R^{j_1,j_2,j_3}_{i_1,i_2,i_3}\cdot \frac{y_1}{y_4}\cdot
\frac{1-\om^{i_2-j_1+1}x_4}{1-\om^{i_2-i_1+1}x_1}$$
which can be rewritten in the form
\beq    R^{j_1,j_2,j_3}_{i_1,i_2,i_3}\;\om^{-j_1}\,=\,
\frac{1}{\om^{i_2+1}x_4}R^{j_1,j_2,j_3}_{i_1,i_2,i_3}
 +\frac{x_1 y_4}{\om^{i_1}y_1 x_4}R^{j_1,j_2,j_3}_{i_1,i_2+1,i_3-1}
  -\frac{y_4}{\om^{i_2+1}x_4 y_1}R^{j_1,j_2,j_3}_{i_1,i_2+1,i_3-1}\ .
\end{equation}
Using the identifications (\ref{xy-uw}) we see that this agrees
with the matrix elements of the first equation of (\ref{matx}).

In order to find the density factor of the modified tetrahedron equation, in
(\ref{rrho}) we have seen that we need the determinant of $\R_{1,2,3}$. Its
calculation is made nontrivial by the presence of the factor
$\delta_{i_2+i_3,j_2+j_3}\;\om^{(j_1-i_1)\,j_3}$ in (\ref{RWdef}). There is
an interesting technology involving the Fermat curve functions $W_p(n)$,
details of which can be found in \cite{Bazh-Bax,SMS,gps},
compare also the appearance
of $W_p(n)$ in connection with the quantum dilogarithm e.g. in eq.(3.7)
of \cite{fk-qd}.
The calculation of $\:\det\:\R\;$is given in detail in \cite{gps}. The
result is:
\beq\label{det-R-fin}
\det\:\R\;=\;N^{N^3}\:\lk\lk\frac{x_4}{y_1\,y_2}\rk^{N(N-1)/2}
\frac{V(x_1)V(x_2)}{V(x_3)V(x_4)} \rk^{\!N^2}\end{equation}
with $\;\:V(x)\;=\;\prod_{\nu=1}^{N-1}\;(1\:-\,\om^{\nu+1}\,x)^\nu.$

\subsection{The matrix $\;R_{i_1,i_2,i_3}^{j_1,j_2,j_3}\;$ for $\:N=2$.}
For $N=2$, using combined indices
$\;i=1+i_1+2 i_2+4 i_3;\;\;j=1+j_1+2 j_2+4 j_3\,\;$ we can give
$\:R_{i_1,i_2,i_3}^{j_1,j_2,j_3}$ explicitly in matrix form:
We define
\bea Y_j&=&\frac{y_j}{1\,+\,x_j}\;=\;\sqrt{\frac{1-x_j}{1+x_j}}
  \hs\mbox{for}\hs j=1,\;2,\;3,\;4\,;\ny\\
  Z_{ij}&=&\frac{Y_i}{Y_j}\,\hq\mbox{for}\hq ij=13,\;14,\;23,\;24;\hs
  Z_{12}=Y_1Y_2;\hs Z_{34}=\frac{1}{Y_3Y_4}.\label{Y2}\eea
and get
\beq \R_i^j=\lk\begin{array}{cc cc cc cc}
1 &\ds Z_{24}&0&0&0&0&\ds Z_{23} &\ds -Z_{34}\\[1mm]
\ds Z_{13}&\ds Z_{13}Z_{24}&0&0&0&0 &\ds-Z_{12} &\ds Z_{14}\\[1mm]
0&0& \ds Z_{13}Z_{24}& \ds Z_{13}& \ds Z_{14}&-Z_{12}&0&0\\[1mm]
0&0& \ds Z_{24}&1& \ds-Z_{34} &\ds Z_{23}&0&0\\[1mm]
0&0& \ds Z_{23} & \ds Z_{34}& 1 &\ds -Z_{24}&0&0\\[1mm]
0&0& \ds Z_{12} &\ds Z_{14} & \ds -Z_{13}&\ds Z_{13}Z_{24}&0&0\\[1mm]
\ds Z_{14}& Z_{12}&0&0&0&0&\ds Z_{13}Z_{24}&\ds -Z_{13}\\[1mm]
\ds Z_{34}&\ds Z_{23}&0&0&0&0&\ds -Z_{24}& 1\\
\end{array}\rk_{ij}\label{Vtm} \end{equation}
The determinant can be calculated directly: \beq
\det\:\R\;=\;\lk Y_1\:Y_2\:(1+Y_3^{-2})(1+Y_4^{-2})\rk^4\,.
\end{equation}

\subsection{The functional mapping in terms of trilinear Hirota equations}
When we wrote the MTE in Sec.\ref{tetrah}, we did not specify the details of
the functional transformations $\Ropf_{i,j,k}$ which act on the scalar
parameters of the quantum operators $\R_{i,j,k}$ (by (\ref{RWdef}) the scalar
parameters then determine the Fermat variables). We are interested to write this
more explicitly and we like to understand which restrictions the MTE imposes on
the $\Ropf_{i,j,k}$.\\[2mm]
Slightly rewritten, eqs. (\ref{lambda-N}), (\ref{funcN}) which define the
functional mapping, are
\bea \frac{{u_3'}^N}{u_3^N}&=&\frac{u_2^N}{{u_2'}^N}=
\lk u_3^{-1}\Lambda_1 u_2\rk^N\;=\;\frac{u_2^N}{u_1^N}+
    \frac{w_1^Nu_2^N}{u_1^Nu_3^N}
   +\ka_1^N\frac{w_1^N}{u_3^N},\ny \\
\frac{u_1^N}{{u_1'}^N}&=&\frac{w_3^N}{{w_3'}^N}=
\lk w_3 \Lambda_2 u_1\rk^N\;=\;\frac{\ka_1^N u_1^N}{\ka_2^N u_2^N}
 +\!\frac{\ka_3^N w_3^N}{\ka_2^N w_2^N}+\!\frac{\ka_1^N
\ka_3^N u_1^N w_3^N}{\ka_2^N u_2^N w_2^N};\ny\\
\frac{{w_1'}^N}{w_1^N}&=&\frac{w_2^N}{{w_2'}^N}=
\lk w_1^{-1}\Lambda_3 w_2\rk^N\;=\;\frac{w_2^N}{w_3^N}+
    \frac{w_2^N u_3^N}{w_1^N w_3^N}+\ka_3^N\frac{u_3^N}{w_1^N}.\label{fun}\eea
If we just use (\ref{fun}) repeatedly to calculate the $\R^{(j)}$ in
(\ref{teab}), we get very lengthy expressions. $\R^{(3)}$ will be
much more complicated than $\R^{(2)}$ etc. It is not transparent, which
variables are independent because the conservation of the centers is hidden
in complicated formulas. However, the nice symmetry of
Fig.\ref{tet-graph} suggests that a different parameterization should be
possible which treats all operators in (\ref{mte-general}) on essentially
equal footing. Indeed this is achieved applying a Legendre transformation
or, equivalently, by the introduction of tau-functions as it is common
practice in the theory of classical integrable systems
\cite{pak-serg-spectr}.
Let us see in an elementary way how this works.

\begin{figure}[t]
\setlength{\unitlength}{0.014mm}
\begin{center}
{\renewcommand{\dashlinestretch}{30}
\begin{picture}(9813,4618)(0,-10)
\path(675,13)(2475,4288)
\path(2456.082,4165.762)(2475.000,4288.000)(2400.784,4189.045)
\path(8458,12)(6658,4287)
\path(6732.216,4188.045)(6658.000,4287.000)(6676.918,4164.762)
\path(5175,463)(9450,2038)
\path(9347.770,1968.365)(9450.000,2038.000)(9327.028,2024.666)
\path(9450,1363)(5400,3163)
\path(5521.842,3141.678)(5400.000,3163.000)(5497.473,3086.849)
\path(3600,1813)(225,3613)
\path(345.000,3583.000)(225.000,3613.000)(316.765,3530.059)
\path(2595,320)(570,3965)
\path(654.502,3874.670)(570.000,3965.000)(602.052,3845.532)
\path(225,688)(3600,2488)
\path(3508.235,2405.059)(3600.000,2488.000)(3480.000,2458.000)
\path(4275,2038)(4950,2038)
\path(4830.000,2008.000)(4950.000,2038.000)(4830.000,2068.000)
\path(5632,74)(7432,4349)
\path(7413.082,4226.762)(7432.000,4349.000)(7357.784,4250.045)
\put(4300,2308){\makebox(0,0)[lb]{${\mathcal{R}_{1,2,3}}$}}
\put(1530,1138){\makebox(0,0)[lb]{$\sk{a_1}$}}
\put(3285,1678){\makebox(0,0)[lb]{$\sk{d_0}$}}
\put(2475,2578){\makebox(0,0)[lb]{$\sk{d_1}$}}
\put(1800,1903){\makebox(0,0)[lb]{$\sk{c_1}$}}
\put(2300,3568){\makebox(0,0)[lb]{$\sk{b_3}$}}
\put(1750,2308){\makebox(0,0)[lb]{$\sk{b_2}$}}
\put(2430,1650){\makebox(0,0)[lb]{$\sk{a_2}$}}
\put(1125,1678){\makebox(0,0)[lb]{$\sk{b_1}$}}
\put(3735,2578){\makebox(0,0)[lb]{$\sk{A}$}}
\put(2520,4423){\makebox(0,0)[lb]{$\sk{B}$}}
\put( 450,4153){\makebox(0,0)[lb]{$\sk{C}$}}
\put(   0,3658){\makebox(0,0)[lb]{$\sk{D}$}}
\put( 810,3703){\makebox(0,0)[lb]{$\sk{c_3}$}}
\put( 360,3208){\makebox(0,0)[lb]{$\sk{d_3}$}}
\put(1440,3000){\makebox(0,0)[lb]{$\sk{d_2}$}}
\put(1080,2578){\makebox(0,0)[lb]{$\sk{c_2}$}}
\put(945,328){\makebox(0,0)[lb]{$\sk{b_0}$}}
\put(495,958){\makebox(0,0)[lb]{$\sk{a_0}$}}
\put(2385,823){\makebox(0,0)[lb]{$\sk{c_0}$}}
\put(3105,2398){\makebox(0,0)[lb]{$\sk{a_3}$}}
\put(8010,328){\makebox(0,0)[lb]{$\sk{c_0}$}}
\put(7965,2173){\makebox(0,0)[lb]{$\sk{d_1}$}}
\put(5805,2713){\makebox(0,0)[lb]{$\sk{d_3}$}}
\put(5130,3208){\makebox(0,0)[lb]{$\sk{D}$}}
\put(6480,4423){\makebox(0,0)[lb]{$\sk{C}$}}
\put(7470,4468){\makebox(0,0)[lb]{$\sk{B}$}}
\put(9675,2083){\makebox(0,0)[lb]{$\sk{A}$}}
\put(9000,2000){\makebox(0,0)[lb]{$\sk{a_3}$}}
\put(9000,1228){\makebox(0,0)[lb]{$\sk{d_0}$}}
\put(8280,1450){\makebox(0,0)[lb]{$\sk{a_2}$}}
\put(6840,880){\makebox(0,0)[lb]{$\sk{a_1}$}}
\put(5400,688){\makebox(0,0)[lb]{$\sk{a_0}$}}
\put(5805,193){\makebox(0,0)[lb]{$\sk{b_0}$}}
\put(7470,1678){\makebox(0,0)[lb]{$\sk{c_1}$}}
\put(6885,2218){\makebox(0,0)[lb]{$\sk{d'_2}$}}
\put(7380,2848){\makebox(0,0)[lb]{$\sk{c'_2}$}}
\put(7290,3703){\makebox(0,0)[lb]{$\sk{b_3}$}}
\put(6615,3793){\makebox(0,0)[lb]{$\sk{c_3}$}}
\put(6660,3073){\makebox(0,0)[lb]{$\sk{b'_2}$}}
\put(1035,3270){\makebox(0,0)[lb]{$\sk{1}$}}
\put(1950,2758){\makebox(0,0)[lb]{$\sk{2}$}}
\put(1395,2083){\makebox(0,0)[lb]{$\sk{3}$}}
\put(3200,2083){\makebox(0,0)[lb]{$\sk{4}$}}
\put(1720,1543){\makebox(0,0)[lb]{$\sk{5}$}}
\put(1040,1183){\makebox(0,0)[lb]{$\sk{6}$}}
\put(8595,1858){\makebox(0,0)[lb]{$\sk{4}$}}
\put(7785,1228){\makebox(0,0)[lb]{$\sk{5}$}}
\put(6435,2443){\makebox(0,0)[lb]{$\sk{2'}$}}
\put(6975,3613){\makebox(0,0)[lb]{$\sk{3'}$}}
\put(7560,2308){\makebox(0,0)[lb]{$\sk{1'}$}}
\put(5805, 823){\makebox(0,0)[lb]{$\sk{6}$}}
\put(6075,1633){\makebox(0,0)[lb]{$\sk{b_1}$}}
\put(1275,4400){\makebox(0,0)[lb]{$Q_1$}}
\put(8575,4400){\makebox(0,0)[lb]{$Q_2$}}
\end{picture}}
\caption{\footnotesize{
Parameterization of the arguments $u_i^{(j)},\;w_i^{(j)}$ of the rational
mapping $\mathcal{R}_{i,j,k}^{(j)}$ in terms of line-section ratios.
}}\label{tau-par}
\end{center}
\end{figure}
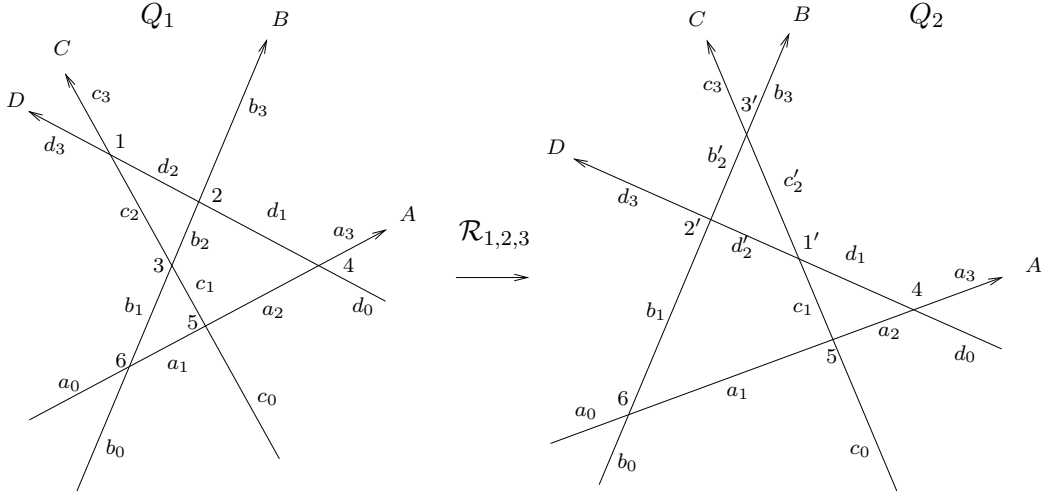

In Fig.\ref{tau-par} we show a magnified view of the two quadrangles $\:Q_1\:$
and $\:Q_2\:$ of Fig.\ref{tet-graph}. Consider $\:Q_1\,$:
The twelve scalar variables $u_i,\:w_i\;\;(i=1,\ldots,6)$ are located at the
six vertices labeled $i=1,\ldots,6$.
The vertices cut each directed line $A,\,B,\,C,\,D$ into four line-sections.
We label these line-sections as shown in the figure. Ratios of line sections
will be our new variables and we explain the rule for the variables
$\,u_1,\:w_1\,$ associated with vertex "1" in the upper left corner:
\begin{itemize}\item
$u_1=c_2/c_3$, i.e. a $u_i$ is expressed as the ratio of the sections
{\it before} to {\it after} point "1" on the {\it right} pointing line through
the vertex "1".
\item $w_1=d_3/d_2$, i.e. a $w_i$ is expressed as the ratio of the sections
{\it after} to {\it before} point "1" on the {\it left} pointing line.
\end{itemize}
This way the twelve scalar variables $u_1,w_1,\ldots,u_6,w_6\:$ of each
quadrangle are expressed in terms
of eight ``internal`` line sections, $\:a_1,a_2,\ldots,d_1,d_2\:$ for $\,Q_1\,$,
and eight ``external`` line sections $\:a_0,a_3,\ldots,$$d_0,d_3\:$.

Passing from one quadrilateral to an adjacent one, corresponding to a mapping
$\Rop$, always three of the ``internal`` lines are changed, and for
distinction, we attach to these changed variables dashes (if left-turning in
Fig.\ref{tet-graph} and daggers (if right turning). The ``external`` line
sections remain unchanged in all mappings and so are no relevant dynamic
variables.

Applied to the transformation $\Ropf_{1,2,3}$ this means that we substitute
\bea u_1=\frac{c_2}{c_3};&\;& u_2=\frac{b_2}{b_3};\hq u_3=\frac{b_1}{b_2};\hq
w_1=\frac{d_3}{d_2};\hq w_2=\frac{d_2}{d_1};\hq w_3=\frac{c_2}{c_1};\ny \\
u_1'=\frac{c_1}{c_2'};&\;& u_2'=\frac{b_1}{b_2'};\hq u_3'=\frac{b_2'}{b_3};\hq
w_1'=\frac{d_2'}{d_1};\hq  w_2'=\frac{d_3}{d_2'};\hq w_3'=\frac{c_3}{c_2'},
 \eea
and (\ref{fun}) become tri-linear equations for the three dashed variables:
\bea {b_2}'^Nc_2^Nd_2^N&=&b_1^Nc_3^Nd_2^N+b_2^Nc_3^Nd_3^N
+\ka_1^Nb_3^Nc_2^Nd_3^N;\ny\\
\ka_2^Nb_2^Nc_2'^Nd_2^N&=&\ka_1^Nb_3^Nc_1^Nd_2^N+\ka_3^Nb_2^Nc_3^Nd_1^N
+\ka_1^N\ka_3^Nb_3^Nc_2^Nd_1^N;\ny\\
b_2^Nc_2^N{d_2}'^N&=&b_2^Nc_1^Nd_3^N+b_1^Nc_1^Nd_2^N
+\ka_3^Nb_1^Nc_2^Nd_1^N;   \label{eqsmot}\eea

The merit of this parameterization is that it automatically
incorporates the invariance of the centers of the mapping:
$\;w_1'w_2'\,=\,w_1w_2\;,\;u_2'u_3'\,=\,u_2u_3\;,\;
u_1'{w_3'}^{-1}\,=\,u_1w_3^{-1}\,.$
It is also simple to express the Fermat variables appearing in (\ref{xy-uw}) in
terms of the line-section parameters. For $\R_{1,2,3}$ we label the
Fermat coordinates by a superscript $^{(1)}$ (as in (\ref{teab})), and get:
\bea x_1^{(1)} &=& \om^{-1/2}\frac{b_2c_3}{\ka_1b_3c_2}\,;\hs
     x_2^{(1)}\;=\;\om^{-1/2}\frac{\ka_2b_1c_2'}{b_2'c_1}\,;\hs\hq\!
     x_3^{(1)}\;=\;\om^{-1}  \frac{b_1c_3}{b_2'c_2}\,;\ny\\
  \frac{y_3^{(1)}}{y_1^{(1)}}&=& \frac{\ka_1b_3d_3}{b_2'd_2}\,;\hs\hs\!\!
  \frac{y_4^{(1)}}{y_1^{(1)}}\;=\;\om^{-1/2}\frac{\ka_3c_2d_1}{c_1d_2}\,;\hs\;
  \frac{y_3^{(1)}}{y_2^{(1)}}\;=\;\frac{c_1d_3}{c_2d_2'}\,.\label{xy-abcd}\eea
Now to find all the functional mappings which appear in the MTE
(\ref{mte-general}), we have to write also the other seven relations
corresponding to $Q_2\ra Q_3\:,\;\ldots,Q_1\ra Q_8$, which look all similar
to (\ref{eqsmot}),~(\ref{xy-abcd}).
E.g. $\R_{1,4,5}$ leading from $Q_2$ to $Q_3$ is governed by $a_2'',\;c_1'',
d_1'', c_2',\;d_2'\;etc.$
which are related by
\bea a_2''^Nc_1^Nd_1^N&=&a_1^Nc_2'^Nd_1^N+a_2^Nc_2'^N{d_2}'^N
+\ka_1^Na_3^Nc_1^N{d_2}'^N;\ny\\
\ka_4^Na_2^Nc_1''^Nd_1^N&=&\ka_1^Na_3^Nc_0^Nd_1^N+\ka_5^Na_2^Nc_2'^Nd_0^N
+\ka_1^N\ka_5^Na_3^Nc_1^Nd_0^N;\ny\\
a_2^Nc_1^Nd_1''^N&=&a_2^Nc_0^N{d_2}'^N+a_1^Nc_0^Nd_1^N
+\ka_5^Na_1^Nc_1^Nd_0^N. \label{eqsmota}\eea
%
Since we know already that $\Ropf_{i,j,k}$
is a solution to the functional TE, one may easily show that eight of the
$\,8\times 3\,$ trilinear eqs. of type (\ref{eqsmot}) are superfluous.
For the whole MTE we have introduced 16 line sections to replace the 12 scalar
variables, but the eight ``external`` variables $\,a_0,\ldots,d_3\,$
and the six couplings $\ka_i$ turn out to be irrelevant, i.e. these can be
eliminated by a re-scaling \cite{gps}.

So the conclusion is that the eight ``internal`` variables
$\,a_1,a_2,\ldots,d_1,d_2\,$ together with sixteen phase
choices parameterize the MTE eq.(\ref{mte-general}) or (\ref{mte-elements}).
This is much more complicated than the rapidities which appear equally on
both sides of a Yang-Baxter-equation. Of course, there is no difference
property of the parameters of the MTE.

\subsection{Solving the MTE using tools of algebraic geometry}

It is well-known \cite{krich,Shi,Mul,KWZ,ser-exact} that discrete
integrable classical equations can be solved by methods of
algebraic geometry. The bilinear Hirota equation is equivalent to
Fay's identity for theta-functions on the Jacobian of an algebraic
curve \cite{KWZ}. The tri-linear equations (\ref{eqsmot}) are also
of the correct type to be solved using Fay's identity.

We use standard concepts, see e.g. \cite{Mum}: Let $\Gamma_g$ be a generic
algebraic curve of genus $g$ and $\om$ the canonical $g$-dimensional vector of
the holomorphic differentials. For any two points $X,\;Y\:\in\,\Gamma_g$
let $\;\:\I:\;\;\Gamma^{\otimes 2}\:\mapsto \mathbb{C}^g$ be the Jacobi
transform:
\beq
\I_X^Y\;\equiv\;\I(X,Y)\;\stackrel{def}{=}\;\int_X^Y\;\om\;\:\in\:
\mathrm{Jac}(\,\Gamma_g\,).    \end{equation}
Let further $E(X,Y)\:=\:-E(Y,X)\:$ be the prime form on $\Gamma^{\otimes 2}$,
$\vj\:\in\:\mathbb{C}^g\,,\,$ and $\;\Theta(\vj)\:$ the zero characteristic
theta-function on $\;\mathrm{Jac}(\,\Gamma_g\,)$.
Then, for $\;A,\:B,\:C\:,D\:\in\,\Gamma_g\;$ Fay's identity states
\bea  \T(\vj)\;\T(\V_B^A\,+\,\I_D^C)&=&
 \T(\V_D^A)\;\T(\V_B^C)\:\frac{E(A,B)\:E(D,C)}{E(A,C)\:E(D,B)}\ny\\ &+&
 \T(\V_B^A)\;\T(\V_D^C)\:\frac{E(A,D)\:E(C,B)}{E(A,C)\:E(D,B)}\:.\label{HI}\eea
In order to solve (\ref{eqsmot}),~(\ref{eqsmota}),$\;\:etc.$ we can build a
tri-linear identity by applying (\ref{HI}) twice:
\bea
\lefteqn{\T(\V_X^Q)\;\T(\V_Y^Q\,+\,\I_Z^{Z'})\;\T(\V_Z^Q\,+\,\I_Y^{Y'})}\ny\\
&=& \T(\V_X^Q\,+\,\I_Z^{Z'})\;\T(\V_Y^Q)\;\T(\V_Z^Q\,+\,\I_Y^{Y'})\:
      \frac{E(Z,X)\:E(Y,Z')}{E(Z',X)\:E(Y,Z)}\ny\\ &+&
 \T(\V_X^Q\,+\,\I_Y^{Y'})\;\T(\V_Z^Q)\;\T(\V_Y^Q\,+\,\I_Z^{Z'})\:
      \frac{E(Y,X)\:E(Y',Z)}{E(Y',X)\:E(Y,Z)}\ny\\ &-&
 \T(\V_X^Q\,+\,\I_Y^{Y'}\,+\,\I_Z^{Z'})\;\T(\V_Y^Q)\;\T(\V_Z^Q)\:
 \frac{E(Z,X)\:E(Y,X)\:E(Y',Z')}{E(Z',X)\:E(Y',X)\:E(Y,Z)}\:.\hx\hq
           \label{HII}\eea
$Q,\:X,\:Y,\:Y',\:Z,\:Z'$ are arbitrary distinct points on $\Gamma_g$.
The dependence on $Q$ is trivial since it appears only in a fixed
combination with $\vj$.
It is not easy to handle the prime forms themselves, but here only cross ratios
of prime forms are needed and these are well-defined quasi-periodical functions
on  $\:\Gamma^{\otimes 4}$:
$$  \frac{E(A,B)\:E(D,C)}{E(A,C)\:E(D,B)}\;=\;
     \frac{\T_{\epsilon_{odd}}(\:\I_A^B\:)\:\T_{\epsilon_{odd}}(\:\I_D^C\:)}
          {\T_{\epsilon_{odd}}(\:\I_A^C\:)\:\T_{\epsilon_{odd}}(\:\I_D^B\:)},$$
where $\T_{\epsilon_{odd}}$ is a non-singular odd characteristic theta function
such that $\T_{\epsilon_{odd}}({\mathbf 0})\:=\,0.$

Consider eqs.(\ref{eqsmot}) for showing the application
of (\ref{HII}) to solve our tri-linear equations.
If we identify
\bea   b_1^N&\sim&\T(\V_X^Q);\hq b_2^N\;\sim\;\T(\V_X^Q\:+\:\I_Y^{Y'});\hq
   b_3^N\;\sim\;\T(\V_X^Q\:+\:\I_Y^{Y'}\:+\:\I_Z^{Z'});\ny \\
   {b_2'}^N&\sim&\T(\V_X^Q\:+\:\I_Z^{Z'});\hq c_2^N\;\sim\;\T(\V_Y^Q);\hq
   c_3^N\;\:\sim\;\T(\V_Y^Q\:+\:\I_Z^{Z'});\ny \\
   d_2^N&\sim&\T(\V_Z^Q\:+\:\I_Y^{Y'});\hq  d_3^N\;\sim\;\T(\V_Z^Q);\hq
   {{d\,}_2'}^N\!\sim\;\T(\V_Z^Q\:+\:\I_X^{X'}). \eea
then, not discussing here the prime form cross ratios, the first eq.
of (\ref{eqsmot}) takes just the form (\ref{HII}). The third eq. of
(\ref{eqsmot}) takes this form if we exchange $X\:\leftrightarrow\:Z\,$ in
(\ref{HII}).
In order to obtain the second eq. of (\ref{eqsmot}) replace in (\ref{HII})
$\;X\:\rightarrow Y',\;Y\:\rightarrow X\;$ and
$\;\vj\:\rightarrow\:\vj+\I_Y^{Y'}$.

Here we shall not give the details \cite{gpsa} of the calculation which shows
that all 24 tri-linear equations can be written in the form (\ref{HII}).
For parameterizing the line section parameters and so parameterizing the eight
$\R^{(j)}$ which appear in the MTE, we have to choose a $\Gamma_g$ and to
introduce just eight points
$X,\:X',\:Y,\:Y',\:Z,\:Z',\:U;\:U'\:\in\:\Gamma_g\:.$

From these formulas simple parameterizations can be obtained
taking the rational limit. Then the theta functions become
trigonometic functions and the cross ratios of the prime forms
become simple rational cross ratios \cite{pak-serg-spectr}.

\section{Free bosonic realization of $\R_{1,2,3}$}

Together with introducing the ZBB-model in \cite{Bazh-Bax-st-tri}, Bazhanov and
Baxter also considered a related continuum bosonic model. Since our framework
includes the ZBB-model, we expect that such a bosonic representation should also
exist for $\R_{1,2,3}$. Since this broadens the class of models, we shall
present here the basic formulas. We shall not consider this realization a
limiting case of the discrete realizations, but rather a new representation.
\\[1mm]
Instead of the cyclic weights
$\;W_p(n)$ of (\ref{W-def}) we introduce the following Gaussian weights:
\beq
W_x(\si)\;=\;\exp{\lk\frac{i}{2\hbar}\:\frac{x}{x-1}\;\si^2\rk}\,;\hx
\hs\sig\in\Rer,\hq x\in{\mathbb{C}};\hq {\Im}m\,\frac{x}{x-1}>0.
\label{Gauss}\end{equation}
At each vertex $j$ of a graph define a pair of operators
$\:\qop_j,\;\pop_j\:$ satisfying
$\:[\,\qop_{j'},\pop_j\:]=i\hbar\delta_{j,j'}$,
and scalar variables $u_j,\;w_j$. We choose a
basis $\;|\,\sig_j\rangle\;$ with $\langle \sig_{j'}|\,\sig_j\rangle=
   \delta(\sig_{j'}-\sig_j),$ such that for $\:\psi(\sig)\in L^2\:$ we have
\beq
\psi(\sig_j)\stackrel{def}{=}
\langle \sig_j|\psi\rangle;\hs
  \langle\sig_j|\:\qop_j\,|\psi\rangle=\sig_j\psi(\sig_j);\hs
  \langle\sig_j|\:\pop_j\,|\psi\rangle=\frac{\hbar}{i}\:
  \frac{\partial\psi(\sig_j)}{\partial\sig_j}. \label{bobas}\end{equation}
For each set of vertices we define the corresponding operators and scalars and
a direct product space of the single vertex spaces.
Consider now the following mapping $\R_{1,2,3}$ in the product space of three
points $\:j=1,2,3$ by
\bea \R_{123}\:\qop_j\:\R_{123}^{-1}&=&\sum_{k=1}^{3}
\frac{\partial \log{u_j'}}
{\partial \log{u_k}}\:\qop_k\;+\;\frac{\partial\log{u_j'}}
{\partial \log{w_k}}\:\pop_k\;\equiv\;\qop_j'\ny\\
\R_{123}\:\pop_j\:\R_{123}^{-1}&=&\sum_{k=1}^{3}
\frac{\partial\log{w_j'}}
{\partial\log{u_k}}\:\qop_k\;+\;\frac{\partial\log{w_j'}}
{\partial\log{w_k}}\:\pop_k\;\equiv\;\pop_j'.\label{trsi}\eea
where the relation between the primed and unprimed scalars $u_j,\;w_j$
is given by (\ref{fctm}) with $\:N=1\,.$ The $\ka_j$ are further parameters
(``coupling constants``) at the vertices $j$.
As the following Proposition shows, equations (\ref{trsi}) are the
bosonic continuum analogs to eqs.(\ref{matx}) of the discrete case.
\begin{prop} In the basis (\ref{bobas}) the operator $\R_{1,2,3}$ has the
following kernel
\beq\label{bR-N-matrix}
\ds\begin{array}{clc} &\ds
\langle\sigma_1^{},\sigma_2^{},\sigma_3^{}| \R_{1,2,3}
|\si'_1,\si'_2,\si'_3\rangle &\\[3mm]
&\ds \hs =\; \delta(\si_2^{}+\si_3^{}-\si'_2-\si'_3)\;
\EXP^{-\frac{i}{\hbar}(\sigma'_1-\sigma^{}_1)\,\sigma'_3}\;
\frac{W_{x_1}(\sigma^{}_2-\sigma^{}_1)\,W_{x_2}(\sigma'_2-\sigma'_1)}
{W_{x_3}(\sigma'_2-\sigma^{}_1)\,W_{x_4}(\sigma^{}_2-\sigma'_1)}\;,
\end{array}\label{bobo}\end{equation}
with the constraint $\; x_1\:x_2\:=\:x_3\:x_4\,.$
In terms of the variables $u_j,\:w_j,\:\ka_j,\:\,(j=1,2,3)$, the
$x_k\;\in\;\Rer\;$ are defined by $\:$
(obtained by putting formally $\om^{-1/2}\ra -1$ in (\ref{xy-uw})):
\beq\label{bx-uw}
x_1^{}\;=\,-\frac{1}{\ka_1}\,\frac{u_2^{}}{u_1^{}}\;;\hq
x_2^{}\;=\,-\ka_2\,\frac{u_2'}{u_1'}\;;\hq
x_3^{}\;=\:\frac{u_2'}{u_1^{}}\;;\hq
x_4^{}\;=\:\frac{\ka_2}{\ka_1}\,\frac{u_2^{}}{u_1'}\,,\end{equation}
where $\;u_1'$ and $u_2'$ are defined as in (\ref{us-f}) with $N=1$.
\end{prop}
\noindent\emph{Proof}:~~
We give the proof for one of the six equations (\ref{trsi}), as the other
equations follow analogously. Let us write shorthand $\;|\sig\rangle\:$ for
$\;|\si_1,\si_2,\si_3\rangle\:$ and $\:d^3\sig\:=\:d\si_1d\si_2d\si_3\:$
etc. We consider:
\beq \int d^3\si'\;\langle\:\si\:|\:\R_{1,2,3}\,|\:\si'\:\rangle
\langle\: \si'\,|\:\qop_3\,|\:\si''\,\rangle\:=\:
\int d^3\si'\;\langle\,\si\,|\,\qop_3'\,|
\,\si'\,\rangle \langle \,\si'\,|\,\R_{1,2,3}\,|\,\si''\rangle
\label{exqd}\end{equation}
which should be satisfied for all
$\;\si_1,\si_2,\si_3,\si_1'',\si_2'',\si_3''$.
Written more explicitly, the kernel (\ref{bR-N-matrix}) is:
$$ R_{\si_1,\si_2,\si_3}^{\si'_1,\si'_2,\si'_3}\;=\;
\delta(\si_2+\si_3-\si'_2-\si'_3)\;
\exp{\lk\frac{i}{2\,\hbar}\:\Sigma(\si,\si')\rk}; $$
where
\bea \Sigma(\si,\si')&=&
\frac{w_1(\si_2-\si_1)^2\,+u_3(\si_1-\si_2')^2}{u_2^{-1}w_1(\ka_1u_1+u_2)}+
   \frac{(\ka_1u_1w_2+\ka_3u_2w_3+\ka_1\ka_3u_1w_3)(\si_2-\si_1')^2}
   {\ka_3w_3(\ka_1u_1+u_2)}\ny\\&&\hspace{-2mm}+\;
 \frac{u_3(\ka_1u_1w_2+\ka_3u_2w_3+\ka_1\ka_3u_1w_3)(\si_2'-\si_1')^2}
 {(\ka_3u_3w_3+w_1w_2+u_3w_2)(\ka_1u_1+u_2)}\; -\,2(\si_1'-\si_1)\si_3'.\eea
From (\ref{trsi}) we get:
$$\qop_3'=\frac{(w_1+u_3)u_2(\qop_2-\qop_1)+u_2u_3\qop_3
          +(u_2+\ka_1u_1)w_1\pop_1}{u_2u_3+u_2w_1+\ka_1u_1w_1}. $$
and (\ref{exqd}) becomes:
\bea \lefteqn{\int d^3\sig'\;\delta(\si_2+\si_3-\si'_2-\si'_3)
\exp{\lk\frac{i}{2\hbar}\Sigma(\sig,\sig')\rk}\;\sig_3''\;
   \delta^3(\si'-\si'')}\ny\\
&=&\int d^3\sig'\;\delta^3(\si-\si')\:
\frac{(w_1+u_3)u_2(\si_2'-\si_1')+u_2u_3\si_3'
  +(u_2+\ka_1u_1)w_1\frac{\hbar}{i}\frac{\partial}{\partial \si_1'}}
 {u_2u_3+u_2w_1+\ka_1u_1w_1}\;\times\ny\\ &&\hspace{1cm}
 \times\:\delta(\si_2'+\si_3'-\si_2''-\si_3'')
 \exp{\lk\frac{i}{2\hbar}\Sigma(\sig',\sig'')\rk}.\label{peq}\eea
Since $$ (u_2+\ka_1u_1)\frac{w_1}{2}\,\frac{\partial\Sigma(\sig',\si'')}
 {\partial\si_1'}\;=\;(w_1+u_3)u_2\sig_1'-u_2w_1\sig_2'-u_2u_3\sig_2''
    +(w_1u_2+\ka_1u_1w_1)\sig_3'',$$  eq.(\ref{peq}) reduces to
$$\si_3''\delta(\si_2+\si_3-\si_2''-\si_3'')=
\frac{u_2u_3(\sig_2+\si_3-\si_2'')+(u_2w_1+\ka_1u_1w_1)\si_3''}
{u_2u_3+u_2w_1+\ka_1u_1w_1}\delta(\si_2+\si_3-\si_2''-\si_3'').$$
\hfill $\Box$\\[2mm]
Certainly, if this bosonic representation is used, the MTE involves
integrations over $\Rer$ instead of summations over ${\mathbb{Z}}_N$. The
bosonic MTE may be proven directly with help of Gaussian integrations.

There are indications that this bosonic model will not be critical all over
and so be physically interesting \cite{ser-phasetr}.

\section{Conclusions}
A large class of integrable 3-dimensional lattice spin models is constructed.
The starting point and central object of the construction is an invertible
canonical automorphism of a triple ultralocal Weyl algebra which satisfies
two physically motivated principles: A current branching rule and a Baxter
Z-invariance. To argue that this automorphism operator fulfils the tetrahedron
equation requires no calculation.
If the Weyl parameter $q$ taken to be a root of unity, one chooses the
$N\times N$ unitary representation of the Weyl operators and the $N$-$th$
powers of the Weyl operators are represented by numbers. Accordingly, in this
representation the automorphism operator can be decomposed into a matrix
conjugation and a functional mapping. Quite trivially, the functional mapping
operator satisfies a tetrahedron equation too. The conjugation matrix
which can be expressed in compact fashion in terms of the Bazhanov-Baxter
Fermat curve cyclic functions, satisfies the modified tetrahedron equation,
(\ref{mte-general}). So we have a quantum problem with coefficients which
are form a classical integrable system themselves. The large class of different
solutions to the classical integrable system leads to a large class of quantum
3D-integrable models. The Zamolodchikov-Baxter-Bazhanov model is obtained if
we choose the trivial classical solution. There is also a Gaussian continuum
representation of the automorphism operator.

Our expectation is that the framework set is broad enough to contain physically
interesting 3D-integrable lattice models. An immediate application is to
perform an asymmetric limit to construct new 2D-integrable chiral models
involving rapidity parameters living on managable algebraic curves. The vast
possibilities to choose convenient parameterization of the 3D-Boltzmann
weights should open new possibilities to solve the 2D-descendent models.
\\[3mm]
{\large\bf Acknowledgements}\\[3mm]
This work was supported in part by the contract INTAS OPEN
00-00055. S.P.'s work was supported in part by the grants
CRDF RM1-2334-MO-02, RFBR 01-01-00539 and the
grant for support of scientific schools
RFBR 00-15-9655. S.S.'s work was supported in part by the
grants CRDF RM1-2334-MO-02 and RFBR 01-01-00201.

\section{Appendix}
\subsection{$\Rop$-mapping from eqs.(\ref{kirch})}
We give some details of the calculation leading from (\ref{kirch}) to
(\ref{Rma}): We use the first four eqs. of (\ref{kirch}) to express the currents
$|\phi_i'\rg$ $(i=1,2,3)$ and $|\phi_2\rg$ in terms of
$|\phi_1\rg$ and $|\phi_3\rg$. This gives the four equations (not commuting any
Weyl variables):
\bea  |\phi_g\rg &\longrightarrow& (\wop_2\wop_1-\wop_1'\wop_2')|\phi_1\rg
  \;+\;(\wop_1'\wop_3-\wop_1'\wop_2'
  +q^{1/2}\wop_2\uop_3-\ka_3\uop_3\wop_3)|\phi_3\rg\;=\;0\ny\\
|\phi_e\rg &\longrightarrow& q^{1/2}(\uop_3'\uop_2'-\uop_2\uop_3)|\phi_3\rg
  \,+\,\lb \ka_1q^{-1/2}\uop_1\wop_1-\!\uop_2\wop_1
  \!-\!q^{1/2}\uop_3'(\uop_1-\!\uop_2')\rb|\phi_1\rg\;=\;0  \ny\\
|\phi_h\rg &\longrightarrow& \hal(\wop_3'\uop_1-\uop_1'\wop_3)
    (|\phi_1\rg-|\phi_3\rg)\ny\\ &&\hx+\;
        \lb \ka_2q^{-1/2}\uop_2'\wop_2'-\uop_1'\wop_2'
-\wop_3'\uop_2'+\hal(\wop_3'\uop_1+\uop_1'\wop_3)\rb(|\phi_1\rg+|\phi_3\rg)
  \;=\;0\ny\\
|\phi_f\rg &\longrightarrow&
  \lb \ka_1\uop_1'\wop_1'\wop_2'-\ka_2\uop_2\wop_2\wop_1
   +q^{1/2}\ka_3\uop_3'\wop_3'(\uop_2'-\uop_1)\rb|\phi_1\rg \ny\\&&\hx+\;
   \lb \ka_1\uop_1'\wop_1'(\wop_2'-\wop_3)\:+q^{1/2}\ka_3\uop_3'\wop_3'\uop_2'
   -q^{1/2}\ka_2\uop_2\wop_2\uop_3\rb\:|\phi_3\rg\;=\;0.\label{kirchi}  \eea
These homogenous eqs. allow the currents $\;|\phi_1\rg\:$ and $\;|\phi_3\rg\:$
to be chosen freely so that their coefficients must vanish and we have
\beq   \wop_1'\wop_2'=\wop_2\wop_1;\hs \uop_3'\uop_2'=\uop_2\uop_3;
    \hs {\wop_3'}^{-1}\uop_1'=\uop_1\wop_3^{-1} \label{cuw}\eeq
We see that for ultralocal Weyl algebras the three products
$\;\wop_1\wop_2,\;\uop_2\uop_3,\;\uop_1\wop_3^{-1}$ are centers of
our mapping. From the second terms of $\;|\phi_g\rg\:$ and
$\;|\phi_e\rg\:$ we get \bea  \wop_1'&=&
\lk\wop_2\wop_1-q^{1/2}\wop_2\uop_3
                           +\ka_3\uop_3\wop_3\rk \wop_3^{-1}\ny\\
\uop_3'&=& \lk\uop_2\uop_3-q^{-1/2}\uop_2\wop_1
                             +\ka_1q^{-1}\uop_1\wop_1\rk\uop_1^{-1}.
\label{ertz} \eea
Using the center eqs.(\ref{cuw}), we can express immediately also $\wop_2'$
and $\uop_2'$ in terms of the unprimed variables. \\
In order to express $\uop_1'$ in terms of the unprimed variables
we take the difference of the two curly brackets in $\:|\phi_f\rg\:$ and
using (\ref{cuw}) obtain
\beq \ka_1\uop_1'\wop_1'\wop_3\:-\:q^{1/2}\ka_3\uop_3'\uop_1'\wop_3\;=\;
    \ka_2\uop_2\wop_2(\wop_1-q^{1/2}\uop_3)   \eeq
Using (\ref{ertz}) and
$\;\uop_i'\,\uop_j'\:=\:q^{\delta_{i,j}}\uop_j'\,\uop_i'\:,
\;$ $\uop_i\,\uop_j\:=\:q^{\delta_{i,j}}\uop_j\,\uop_i\;$ leads to
\beq \uop_1'\lb \ka_1\wop_2\:+\:q\ka_3\uop_2\uop_1^{-1}\wop_3\:
   +q^{1/2}\ka_1\ka_3\wop_3\rb(\wop_1-q^{1/2}\uop_3)\;=
   \;\ka_2\uop_2\wop_2(\wop_1-q^{1/2}\uop_3).\label{zwes}  \eeq
One easily checks that (\ref{cuw}),~(\ref{ertz}),~(\ref{zwes}) can be written
as (\ref{Rma}),\,(\ref{Lam}).
The inverse relations look similar, we give only one term: To express
 $\uop_1$ or $\wop_3$ in terms of primed variables, we use the second term of
$|\phi_h\rg$, obtaining immediately
\beq \uop_1\;=\;\lk\uop_1'\wop_2'+\wop_3'\uop_2'-\ka_2q^{-1/2}
        \uop_2'\wop_2'\rk\:{\wop_3'}^{-1}.  \eeq
\subsection{$\Rop$-mapping from ``Linear Problem``}

It is often useful, instead of requiring (\ref{kirch}), to start
the derivation of the mapping $\Rop_{1,2,3}$ postulating six
"linear problem"-equations for "co-currents" $\Lg\phi_i|$.

One may understand the nature of the mapping and the origin of the
local Weyl algebra relations if we start from the most general
consideration, i.e. introducing at each vertex {\it three} dynamic variables.

In each sector $\phi_j$ of the auxiliary plane there shall be a
co-current $\Lg\phi_j|$. At each vertex $i$ some dynamic elements
$\uop_i$, $\wop_i$ and $\vop_i$ acts on the
four co-currents of the surrounding sectors such that \\[2mm]
1) the interaction with the co-current $\Lg\phi_j|$ of the sector {\it between}
the arrows leads to a co-current $\:\Lg\phi_j|\:q^{1/2}\uop_i\:$ at the vertex;
with a co-current $\Lg\phi_j|$ in the sector to the {\it left}
of the arrows: to $\:\Lg\phi_j|\:.$  With the $\Lg\phi_j|$ of the sector to
the {\it right} of the arrows the interaction produces
$\:\Lg\phi_j|\,\vop_i\:,\;\:etc.$ \\[2mm]
2) The total co-current at a vertex is obtained by linear superposition of
the four contributions and it must vanish,\\[2mm]
3) These properties are unchanged if one line is shifted through the vertex $i$
($\nabla-\Delta$-invariance).\\[3mm]
The dynamic elements $\uop_i$, $\wop_i$ and $\vop_i$ form an
associative noncommutative ring (actually a body). Co-currents
belong to the formal left module of this ring. Note that the
co-currents are not the conjugates of the currents used in
eqs.$(\ref{kirch})$.\\[1mm]
Applied to the six vertices appearing in
Fig.\ref{fig-YBE}, these rules lead to the following six equations: \bea
\Lg\phi_1|&\equiv&\Lg\phi_c|\;
+\;\Lg\phi_d|\:q^{1/2}\uop_1\;+\;\Lg\phi_a|\:\wop_1\;
+\;\Lg\phi_e|\:\vop_1\;=\;0\:,\ny\\[2mm]
\Lg\phi_2|&\equiv&\Lg\phi_a|\;
+\;\Lg\phi_e|\:q^{1/2}\uop_2\;+\;\Lg\phi_g|\:\wop_2\;
+\;\Lg\phi_f|\:\vop_2\;=\;0\:,\ny\\[2mm]
\Lg\phi_3|&\equiv&\Lg\phi_c|\;
+\;\Lg\phi_a|\:q^{1/2}\uop_3\;+\;\Lg\phi_b|\:\wop_3\;
+\;\Lg\phi_g|\:\vop_3\;=\;0\:,\ny\\[2mm]
\Lg\phi_1'|&\equiv&\Lg\phi_b|\;
+\;\Lg\phi_h|\:q^{1/2}\uop_1'\;+\;\Lg\phi_g|\:\wop_1'\;
+\;\Lg\phi_f|\:\vop_1'\;=\;0\:,\ny\\[2mm]
\Lg\phi_2'|&\equiv&\Lg\phi_c|\;
+\;\Lg\phi_d|\:q^{1/2}\uop_2'\;+\;\Lg\phi_b|\:\wop_2'\;\,
+\;\Lg\phi_h|\:\vop_2'\;=\;0\:,\ny\\[2mm]
\Lg\phi_3'|&\equiv&\Lg\phi_d|\;
+\;\Lg\phi_e|\:q^{1/2}\uop_3'\;+\;\Lg\phi_h|\:\wop_3'\;
+\;\Lg\phi_f|\:\vop_3'\;=\;0\:.\label{linearp}\eea Here
$q^{1/2}\in\mathbb{C}$ is just a scale factor and plays no role.
The mapping is provided by the demand of the linear equivalence of
$\Delta$-shape  and $\nabla$-shape triangles: any two equations of
(\ref{linearp}) must be linear combinations of the rest four. In
the other words, any four external co-currents are to be linearly
independent.

As an example of the application of this principle consider two
linear combinations which are designed such that both do not contain
the co-currents
$\Lg\phi_a|,\;\Lg\phi_c|,\;\Lg\phi_h|$: \bea \Lg\psi'| &\equiv&
\Lg\phi_1'|-\Lg\phi_3'|(\wop_3')^{-1}q^{1/2}\uop_1'\;,\ny\\[2mm]
\Lg\psi| &\equiv& \Lg\phi_3|\wop_3^{-1}-\Lg\phi_1|\wop_3^{-1}+
\Lg\phi_2|(\wop_1^{}-q^{1/2}\uop_3^{})\wop_3^{-1}\;. \eea
Explicitly, using (\ref{linearp}), we have \bea \Lg\psi'| &=&
\Lg\phi_b|+\Lg\phi_g|\wop_1'-\Lg\phi_d|\wop_3^{\prime-1}q^{1/2}\uop_1'\ny\\[2mm]
&&-\Lg\phi_e|q\uop_3'\wop_3^{\prime-1}\uop_1'+
\Lg\phi_f|(\vop_1'-\vop_3'\wop_3^{\prime-1}q^{1/2}\uop_1')\:,\\[2mm]
\Lg\psi| &=&
\Lg\phi_b|+\Lg\phi_g|
(\vop_3^{}\wop_3^{-1}+\wop_2^{}\wop_1^{}\wop_3^{-1}-
q^{1/2}\wop_2^{}\uop_3^{}\wop_3^{-1})\ny\\[2mm]
&&-\Lg\phi_d|q^{1/2}\uop_1^{}\wop_3^{-1}-\Lg\phi_e|
(\vop_1^{}\wop_3^{-1}+q\uop_2^{}\uop_3^{}\wop_3^{-1}-
q^{1/2}\uop_2^{}\wop_1^{}\wop_3^{-1})\ny\\[2mm]
&&+\Lg\phi_f|\vop_2^{}(\wop_1^{}-q^{1/2}\uop_3^{})\wop_3^{-1}\;.
\eea The difference $\Lg\psi'|-\Lg\psi|$ contains exactly four
external co-currents, therefore due to the $\textrm{rank}=4$
demand all the coefficients in $\Lg\psi'|-\Lg\psi|$ must vanish
identically. This gives in particular the expressions for
$\wop_1'$ and $\uop_3'$ in the terms of the unprimed elements, and
some other relations.

In the same way one may consider all the other combinations of
$\Lg\phi_i|\,,\,\ldots\,,\,\Lg\phi_3'|\:$ giving linear combinations of four
external co-currents. There are exactly eight independent
relations for $\uop_i\,,\,\ldots\,,\,\vop_i'\,$. The solution of the linear
equivalence problem may be written as follows: First, the six
primed elements may be expressed explicitly
\begin{equation}
\begin{array}{lll}
\wop_1'=\wop_2^{}\Lambda_3^{}\;,& \wop_2^{\prime
-1}=\wop_1^{-1}\Lambda_3^{}\;,\;\;\;&
\uop_3'\vop_3^{\prime-1}=\vop_1^{}\wop_1^{-1}\widetilde{\Lambda_2}\;,\\
\wop_1'\vop_1^{\prime-1}=\vop_3^{}\uop_3^{-1}\widetilde{\Lambda_2}\;,\;\;\;&
\uop_2^{\prime-1}=\uop_3^{-1}\Lambda_1^{}\;,\;\;\;&
\uop_3'=\uop_2^{}\Lambda_1^{}\;,\end{array} \label{6expr}
\end{equation}
where \bea \Lambda_1 &=&
\uop_3^{}\uop_1^{-1}-q^{-1/2}\wop_1^{}\uop_1^{-1}
                  +q^{-1}\uop_2^{-1}\vop_1^{}\uop_1^{-1}\;,\ny\\[2mm]
\widetilde{\Lambda_2} &=&
\uop_3^{}\vop_3^{-1}\wop_2^{}\vop_2^{-1}+
\wop_1^{}\vop_1^{-1}\uop_2^{}\vop_2^{-1}-q^{-1/2}\vop_2^{-1}\;,\ny\\[2mm]
\Lambda_3 &=& \wop_1^{}\wop_3^{-1}-q^{1/2}\uop_3^{}\wop_3^{-1}+
\wop_2^{-1}\vop_3^{}\wop_3^{-1}\;.\label{3L} \eea Also there exist
three extra relations written below in the left column
\begin{equation}
\begin{array}{lll}
\wop_3^{\prime-1}\uop_1'=\uop_1^{}\wop_3^{-1} &\hs\hq\textrm{\small vs
}\hs\hq& \uop_3'\vop_3^{\prime-1}\vop_1'\wop_1^{\prime-1} =
\vop_1^{}\wop_1^{-1}\uop_3^{}\vop_3^{-1},\\
\vop_1'\uop_1^{\prime-1}\vop_2'\uop_2^{\prime-1}
                       =\vop_2^{}\uop_2^{-1}\vop_1^{}\uop_1^{-1}
&\hs\hq\textrm{\small vs }\hs\hq& \wop_1'\wop_2'=\wop_2^{}\wop_1^{},\\
\vop_3'\wop_3^{\prime-1}\vop_2'\wop_2^{\prime-1}
                       =\vop_2^{}\wop_2^{-1}\vop_3^{}\wop_3^{-1}
&\hs\hq\textrm{\small vs }\hs\hq& \uop_3'\uop_2'=\uop_2^{}\uop_3^{}.
\end{array}\label{6cons}
\end{equation}
The right column in (\ref{6cons}) is just an evident consequence of
(\ref{6expr}).

Two important notes are to be made. Recall, at the first,
expressions (\ref{6expr}),(\ref{3L}),(\ref{6cons}) are obtained in the
framework of the associative noncommutative ring, i.e. we never changed the
order of the invertible elements $\uop_i$, $\wop_i$, $\vop_i$,
$\uop_i'$, $\wop_i'$ and $\vop_i'$. Secondly, since the linear
equivalence principle implies eight independent relations, there
are eight independent relations between nine relations
(\ref{6expr}),(\ref{6cons}). In the framework of the noncommutative
ring the proof of this miracle is a quite complicated exercise.

Now the question is how to get rid of one degree of freedom in the
primed elements. The answer lies in the comparison of the left and
right columns of (\ref{6cons}). One may see that it is natural to
identify left and right columns. The identification produces two
demands
\begin{equation}\label{uwv-id}
\vop_i^{}\uop_i^{-1}\sim\wop_i^{}\;,\hs \vop_i^{}\wop_i^{-1}\sim\uop_i^{}
\end{equation}
and the analogous relations for the primed elements. Evidently, eqs.
(\ref{uwv-id}) fix the Weyl algebra structure.

Let now the initial elements $\uop_i$, $\wop_i$ and $\vop_i$ obey
the local Weyl algebra relations (\ref{Weyl}) with
$\vop_i=\ka_i\uop_i\wop_i$, $\ka_i\in\mathbb{C}$. Then
(\ref{6expr}),(\ref{6cons}) immediately give
$\vop_i'=\ka_i'\uop_i'\wop_i'$, where $\ka_i'$ are centers
such that $\ka_1^{}\ka_2^{}=\ka_1'\ka_2'$,
$\ka_2^{}\ka_3^{}=\ka_2'\ka_3'$, and the mapping (\ref{6expr})
defines the automorphism of the local Weyl algebra. External
``physical'' demand is that since $\ka_i$ are centers, a well
defined automorphism should not change them: $\ka_i'=\ka_i^{}$.
After the identification of $\ka_i$ and $\ka_i'$, eqs.
(\ref{6expr}),(\ref{3L}) coincide with (\ref{Rma}),(\ref{Lam}) up to
the rescaling $\;\Lambda_2=q\ka_1\ka_3\widetilde{\Lambda_2}$.

\bibliographystyle{amsplain}

\end{document}